\documentclass[12pt]{iopart}
\usepackage{iopams} 
\usepackage{graphicx}
\usepackage{adjustbox} 
\usepackage{multirow} 
\usepackage{url}
\usepackage{setspace} 
\begin{document}

\title[Two-body problem in curved spacetime]{Two-body problem in curved spacetime: exploring 
gravitational wave transient cases}

\author{V. N. Yershov$^1$\footnote{Present address:
7 Barley Close, Crawley RH10 6BA, UK.}, A. A. Raikov$^2$ and E. A. Popova$^3$}

\address{$^1$Formerly: Mullard Space Science Laboratory, University College London}
\address{$^2$Special Astrophysical Observatory,  Niznii Arkhyz 369167, Russia}
\address{$^3$Pulkovo Observatory,  Pulkovskoye shosse 65 (1), St. Petersburg 196140, Russia
}
\ead{vyershov@moniteye.co.uk}

\begin{spacing}{0.8} 
\begin{abstract}
Gravitational-Wave Transient Catalogues (GWTC) from the LIGO-Virgo-KAGRA 
collaborations (LVC and LVK) contain almost a hundred gravitational wave (GW) detection cases.
We explore them from the perspective of the two-body problem in curved spacetime, starting 
with the first case, GW150914, which marks the GW discovery \cite{abbott16}.  
In this paper, the LVC authors  estimated the characteristic (chirp) mass 
of the binary blackhole system emitted this signal. Their calculation was based on  
Numerical-Relativity (NR) templates and presumably accounted fully for  
the non-linearity of GR.  
The same team later presented an alternative analysis of GW150914 \cite{abbott17}, 
using the quadrupole post-Newtonian (PN) approximation of GR.
Both analyses gave similar results, despite being based on quite different assumptions about 
the linearity or non-linearity of the coordinate reference frame near the GW source.

Here we revisit the PN-analysis of GW150914 for which we use 
less noisy input GW frequencies, as we have filtered them by reading them from the 
time-frequency map of GW150914. As in paper \cite{abbott17}, our result also agrees with the NR-based chirp 
mass value published in \cite{abbott16}.
Additionally,  we apply the PN-approximation formalism to the rest of the GWTC cases, finding that 
practically all of their PN-approximated chirp masses coincide with the published NR-based values from GWTC.   
In our view, this  implies that the NR-based theory, which is currently in use for processing GW signals,   
does not fully account for the difference between the source and detector 
reference frames because the PN-approximation, which is used for the comparison, 
does not account for this difference by design, given the flat-spacetime initial assumptions 
of this approximation.  

We find that the basis of this issue lies in the source-to-detector coordinate 
transformation. For example, when obtaining the equation of motion of a coalescing binary system by
integrating its energy-momentum tensor and varying the corresponding reduced action functional,
the lapse and shift functions are not involved within the Arnowitt-Deser-Misner
(ADM) parametrisation scheme, which is typically used for the NR-based calculation of GW waveforms. 
A similar non-involvement of the lapse and shift functions is known to occur in the 
description of motion of an orbiter around a Schwarzschild blackhole.
Here the GR expression for the orbital angular frequency, as seen by a remote observer, 
coincides with the Keplerian non-relativistic formula until the very last orbits 
before the plunge phase (although being fully GR-compliant). 
This non-involvement of the time lapse function renders the source-to-detector coordinate 
transformation suitable for building GW waveforms corresponding to the detector frame.
However, the inverse (detector-to-source) transformation requires the derivatives of GW frequencies to be known 
in the source reference frame.  The lack of this knowledge leads to a systematic error
in the estimated chirp masses of GW sources. The corresponding luminosity distances 
of these sources also turn out to be overestimated.
\end{abstract}
\end{spacing}
%
\noindent{\it Keywords}: compact binaries,  black hole mergers, relativistic reference frames,  gravitational waves.
%
%
%
%

\section{Introduction}

After the first three observing runs of the LIGO/Virgo and LIGO/Virgo/KAGRA 
collaborations (LVC and LVK), the number of gravitational wave (GW) 
detection cases has amounted to almost a hundred, which allows us to explore
the statistical properties of relativistic compact objects 
(blackholes and neutron stars) discovered by gravitational wave observatories.    
The parameters of these objects are summarised in a series of Gravitational-Wave 
Transient Catalogues (GWTC) published by LVC, LVK and some other research groups  
\cite{abbott19,abbott20,venumadhav20,abbott21a, abbott21, nitz20}. 

The statistical properties of blackholes found in GW transients  
are quite different from those previously discovered in X-ray binary systems
from our Galaxy.   
This led to a broad discussion about the origin of blackhole binary systems, which  
was prompted by the fact that the masses of blackholes discovered via  
GW detections are very high, often exceeding the upper limit predicted  by the 
modern theory of stellar evolution. Theoretically \cite{fryer01,raithel18}, most 
of the blackholes formed at the end point of stellar evolution  
are predicted to have masses from 10$\,M_\odot$ to $12\,M_\odot$, which was confirmed by
the observed distribution of blackhole masses in Galactic  
X-ray binaries \cite{wiktorowicz13,corralsantana16}.
By contrast, the majority of blackhole masses 
inferred from the GW detections are between 20$\,M_\odot$ and $50\,M_\odot$,  
reaching $\sim100\,M_\odot$ in the cases of GW190521 \cite{abbott20} and GW190426 \cite{abbott21a}.

The first GW signal detected by LVC (GW150914)
was attributed to the coalescence of two  blackholes with  nearly-equal masses 
of $36^{+5}_{-4}$\,M$_\odot$ and $29^{+4}_{-4}$\,M$_\odot$ \cite{abbott16}.
These masses were determined using numerical relativity (NR) waveform templates, 
prepared by taking into account the full non-linearity of General Relativity (GR).  

The same LVC authors published yet another analysis of the same GW signal 
\cite{abbott17}, where they estimated the parameters of this coalescing binary blackhole 
system via an alternative approach -- by using the quadrupole  post-Newtonian (PN) approximation of 
General Relativity. The authors have shown that the PN-approximation leads to almost the same 
result as was obtained in their previous work based on rigorous calculations. 
For example, the calculation based on the PN-approximation formalism applied to the detected
frequencies of the GW signal gives the binary component masses of about 35M$_\odot$ \cite{abbott17}, 
which is similar to the outcome of the NR-template processing scheme in \cite{abbott16}:
$m_1=36^{+5}_{-4}$\,M$_\odot$ and $m_2 = 29^{+4}_{-4}$\,M$_\odot$.

In our view, this agreement between the outcomes of the fully general-relativistic analysis
and of the PN approximation needs some focused attention
because the PN approximation is built upon the assumption of flat spacetime 
and slow motion of objects \cite{einstein16,einstein18,landau60, peters63,peters64},
whereas the full GR-approach deals  with coordinate reference 
frames which are strongly distorted by non-linearities when the distances between two 
blackholes are small.
The unusual effectiveness of the PN approximation when dealing with  highly-relativistic and fast-moving 
binary systems was noticed by C.M. Will \cite{will11}, who wrote that  ``the reasons for this
effectiveness are largely unknown''.

Besides the analysis of the gravitational wave signal GW150914
by LVC \cite{abbott17}, there are  other works 
discussing properties of this signal from different points of view
and with different purposes: see 
\cite{rodriguez16,burko17,mathur17,thorne18,jahanvi19,becsy20,brown21}.
Here we shall return to the analysis of the same signal, with the purpose  
of clarifying the question posed by C.M. Will about how the linearised PN approximation of 
General Relativity (i.e. the approximation of flat spacetime) could characterise 
so well the motion of relativistic objects 
with large masses (exceeding 30\,M$_\odot$) whilst located so close to each other that 
one can hardly neglect the distortions related to their local reference frame. 

We shall also revisit the rest of the GW transient cases from GWTC, applying to them 
the same PN-approximation formalism and comparing the resulting PN-based chirp masses with those obtained rigorously 
and published in GWTC. 

The result of our analysis is surprising. We find that there exist a systematic error 
in both rigorous and PN-based methods of determining the characteristic (chirp) masses of 
coalescing binary systems. 
Here we shall limit our analysis to just the necessary (say, first-order) calculations, 
as our purpose is to present no more than a proof of existence of this systematic error.


\section{PN approximation}

As we are going to apply the PN formalism to the gravitational wave transient cases 
published in GWTC, we need to briefly reproduce here the main points of this formalism. 
Despite the PN  is well-known and well-described in numerous publications, including one of the 
present authors, the systematic error issue which we have mentioned above in \S1 is quite subtle. 
This subtlety requires certain logical steps in the explanation of the origin of this issue, 
which must be followed by the interested reader to see at first hand that the aforementioned 
systematic error does really exist.  

We shall start with the first linearisation of General Relativity that led 
Einstein to his theoretical discovery of gravitational waves \cite{einstein16,einstein18}.
He presented the metric tensor in the form of expansion 
\vspace{-0.3cm}
\begin{equation*}
 g_{\mu\nu} = - \delta_{\mu\nu} + \gamma_{\mu\nu},
\end{equation*}
\noindent
where $\delta_{\mu\nu}$ is the Minkowski flat metric, and $\gamma_{\mu\nu}$ is
a small perturbation which can propagate as a wave (nowadays it is usually denoted 
as $h_{\mu\nu}$ or $h_{ij}$). 
The first attempts to solve the N-body problem by using the PN 
expansion of the GR equations was made by  W. de Sitter 
\cite{desitter16}, as well as by H. A. Lorentz and J. Droste \cite{lorentz17}, which 
like Einstein's method, used a linearised spacetime metric. Approximately at the same time, 
A. Einstein computed the energy emitted in the form of gravitational waves 
by an elongated body, such as a rod or dumbbell \cite{einstein16}. 
This quadrupole approximation was further developed by L.D. Landau and E.M. Lifshitz \cite{landau60}.
It allows  one to calculate the loss of energy of a binary system due to the radiation
of gravitational waves. In 1963, P.C. Peters and J. Mathews \cite{peters63,peters64} used the Einstein-Landau-Lifshitz 
formalism, to devise a solution  allowing the prediction of the evolution 
of orbital parameters of relativistic binary systems, which was later used by various researchers 
for determining orbits of binary pulsars. The first of such systems was a binary pulsar 
PSR B1913+16 \cite{hulse75}, whose orbital parameters were found to evolve in full agreement with 
Peters and Mathews' theoretical prediction. This agreement between theory and 
observations was interpreted as very strong, albeit indirect, evidence of the 
existence of gravitational waves. 
Nowadays, the GW waveforms are calculated by applying higher PN orders \cite{blanchet14,khalil20,blumlein20}
and by using some more advanced techniques, such as the effective one-body (EOB)
formalism \cite{buonanno99} and NR simulations \cite{kidder00,pretorius05,boyle19,ossokine20}.

The basis of the PN approximation is the condition that the gravity fields are weak 
inside and outside  gravitating bodies, that distances under consideration are 
large compared to the size of the studied system, and that matter in the system 
moves with small velocities $v$ as compared to the speed of light $c$, which means that
the system can be described by using a small parameter 
\begin{equation}
\varepsilon \sim (v/c)^2,
\label{epsilon}
\end{equation}
The Einstein-Landau-Lifshitz quadrupole moment formalism for
a post-Newtonian source is described in numerous reviews, see, e.g., 
\cite{will11, blanchet14,arimoto21,gurfil22}. Here we shall follow the description 
by the LVC collaboration \cite{abbott17}, emphasising the basic physics 
of a binary blackhole merger. Using the approach by Landau \& Lifshitz \cite{landau60}, 
the LVC authors define the tensor of the quadrupole moment $Q_{ij}$  as
\begin{equation}
Q_{ij}= \int \mathrm{d}^3x \rho(\mathbf{x)} \left( x_i x_j - \frac{r^2 \delta_{ij}}{3} \right),
\label{pn_quad_lvc}
\end{equation}
where $\mathbf{x}=(x,y,z)$ are Cartesian coordinates, $\rho(\mathbf{x})$ is the mass density,
and $\delta_{ij}= \mathrm{diag}(1,1,1)$ is the Kronecker-delta.
For two bodies $a=\{1,2\}$ moving along a circular orbit 
at separation $r$ with the orbital frequency $\omega_\mathrm{orb}$
\begin{equation}
Q_{ij}^a(t)= \frac{m_a r_a^2}{2} I_{ij},
\label{pn_quad_lvc2}
\end{equation}
where 
\begin{eqnarray}
I_{xx}&=&\cos(2 \omega_\mathrm{orb} t)+ \frac{1}{3}\,,\,\nonumber\\ 
I_{yy}\;&=&\;\frac{1}{3} - \cos(2 \omega_\mathrm{orb} t),\nonumber\\
I_{xy}&=&I_{yx}= \sin (2\omega_\mathrm{orb} t)\,,\nonumber\\
I_{zz}\;&=&\; -\frac{2}{3}.
\label{quad_inertia_moments}
\end{eqnarray}
%
The energy carried away from the system by the radiated gravitational waves is 

\begin{eqnarray}
\frac{\mathrm{d}}{\mathrm{d}t} E_{\mathrm{gw}} &= \frac{1}{5} \frac{G}{c^5} \sum_{i,j=1}^3 
\frac{\mathrm{d}^3Q_{ij}}{\mathrm{d}t^3} \frac{\mathrm{d}^3Q_{ij}}{\mathrm{d}t^3}\nonumber\\ &= 
\frac{32}{5} \frac{G}{c^5} \mu^2 r^4 \omega_{\tt orb}^6,
\label{pn_energy_lvc2}
\end{eqnarray}
where $\mu = \frac{m_1 m_2}{m_1+m_2} $ is the reduced mass
in the equivalent one-body problem (the equivalent one-body problem is the common
standard way of dealing with two-body systems via the mapping of the relative motion
of two components of the system to the motion of a single mass $\mu$ in the potential of a fictitious 
central mass $M=m_1 +m_2$, see, e.g. \cite{hilditch01}, chapter 2, or \cite{blinder04}, chapter 7).    
So, the Keplerian orbital energy 
\begin{equation}
E_\mathrm{orb}= - \frac{GM\mu}{2r}
\label{pn_keplerian_orb_lvc2}
\end{equation}
diminishes by the amount
\begin{equation}
\frac{\mathrm{d}}{\mathrm{d}t} E_{\mathrm{orb}}= 
-\frac{\mathrm{d}}{\mathrm{d}t} E_{\mathrm{gw}}.
\label{pn_de2dt_equalgw}
\end{equation}
The time derivative of (\ref{pn_keplerian_orb_lvc2})  
\begin{equation}
\frac{\mathrm{d}}{\mathrm{d}t} E_{\mathrm{orb}}=\frac{GM\mu}{2r^2}\mathbf{\dot{r}}
\label{pn_keplerian_derivative_lvc2}
\end{equation} 
can be related to (\ref{pn_energy_lvc2}) via (\ref{pn_de2dt_equalgw}) 
with the purpose to calculate parameters of a coalescing binary system.  
In these considerations each orbit is described as approximately Keplerian
(we shall return to this point in \S4).

Using Kepler's third law $r^3 = \frac{GM}{\omega^2}$ and its time derivative 
$\dot{r} = - \frac{2}{3} \frac{r \dot{\omega}}{\omega}$ the LVC authors 
calculate the orbital frequency growth due to the emission of gravitational waves:
\begin{eqnarray}
\dot{\omega}^3_\mathrm{orb}&=& \left(\frac{96}{5}\right)^3 \frac{\omega^{11}_\mathrm{orb}}{c^{15}} G^5 \mu^3 M^2\nonumber\\
&=& \left(\frac{96}{5}\right)^3 \frac{\omega^{11}_\mathrm{orb}}{c^{15}} (G \mathcal{M})^5, 
\label{pn_orbenergy_lvc}
\end{eqnarray}
where
\begin{equation}
\mathcal{M} = (\mu^3 M^2)^\frac{1}{5} = \frac{(m_1 m_2)^{3/5}}{(m_1 +m_2)^{1/5}} 
 \label{chirp_mass}
\end{equation}
is the characteristic (chirp) mass 
of the system.
Thus, by knowing two parameters, the frequency 
of a gravitational wave signal $f_\mathrm{gw}= 2\,f_\mathrm{orb} = \omega_\mathrm{orb}/\pi$ and its 
time derivative $\dot{f}_\mathrm{gw}$, 
one can find this mass according to the formula  
{\small
\begin{equation}
 \mathcal{M}(f_\mathrm{gw}, \dot{f}_\mathrm{gw})=  
 \frac{c^3}{G}  \Big[ \frac{5}{96} \pi^{-8/3} f^{-11/3}_\mathrm{gw} \dot{f}_\mathrm{gw} \Big]^{3/5}.
 \label{freq2chirp_mass}
\end{equation}
}
\noindent
The frequency of the GW signal is $f_\mathrm{gw}$ is twice the orbital frequency is explained by the fact
that the leading term of the multipole expansion in the case of a two-body system is quadrupole.
Therefore, there is neither monopole nor dipole radiation, and the components of the moment of inertia 
(\ref{quad_inertia_moments})  contain only the double orbital frequency terms $2 \omega_{\mathrm orb}$
(for more details, see, e.g., \S3.3 in \cite{maggiore08}).   

Let us analyse the gravitational-wave
signal GW150914 using formula (\ref{freq2chirp_mass}). 

\section{Gravitational-wave signal GW150914}

\begin{figure}[htb]
\vspace{-0.6cm}
\hspace{0.6cm}
\includegraphics[scale=0.4]{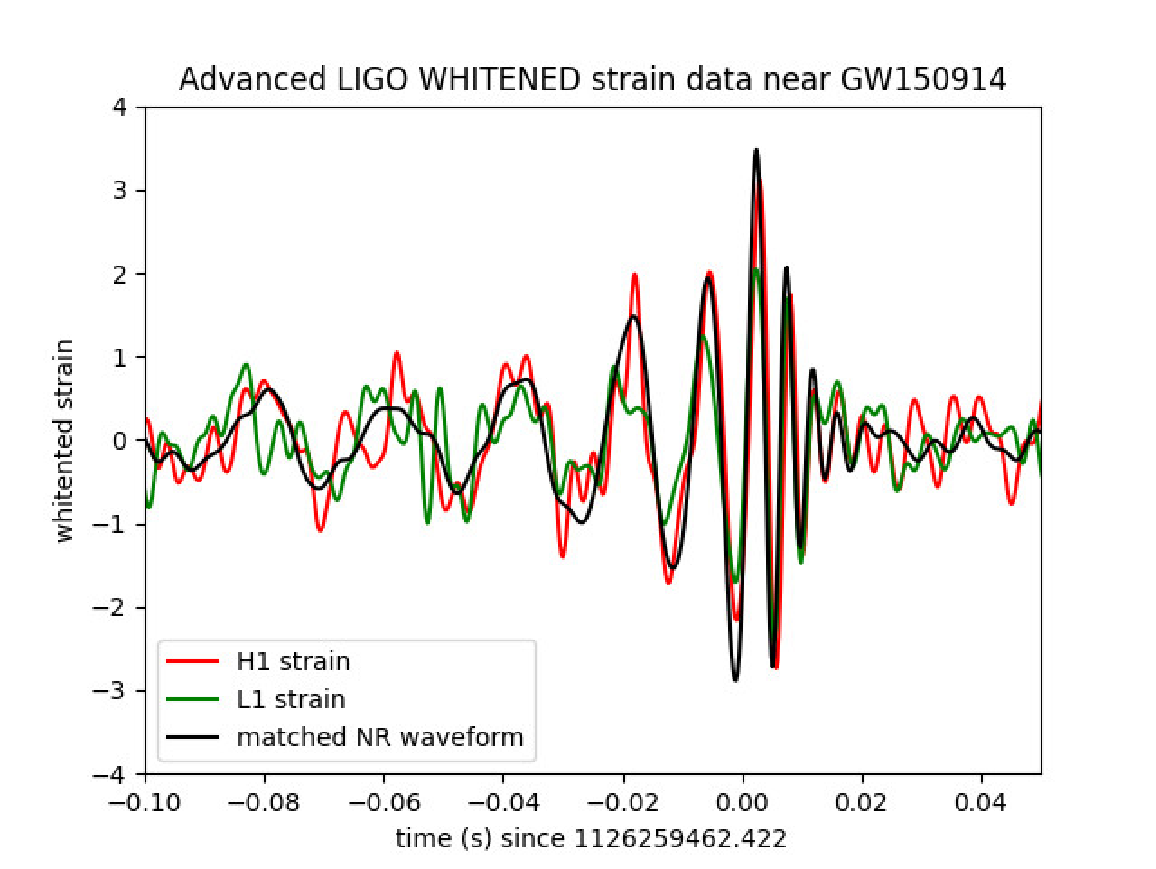}
\hspace{-0.1cm}
\includegraphics[scale=0.35]{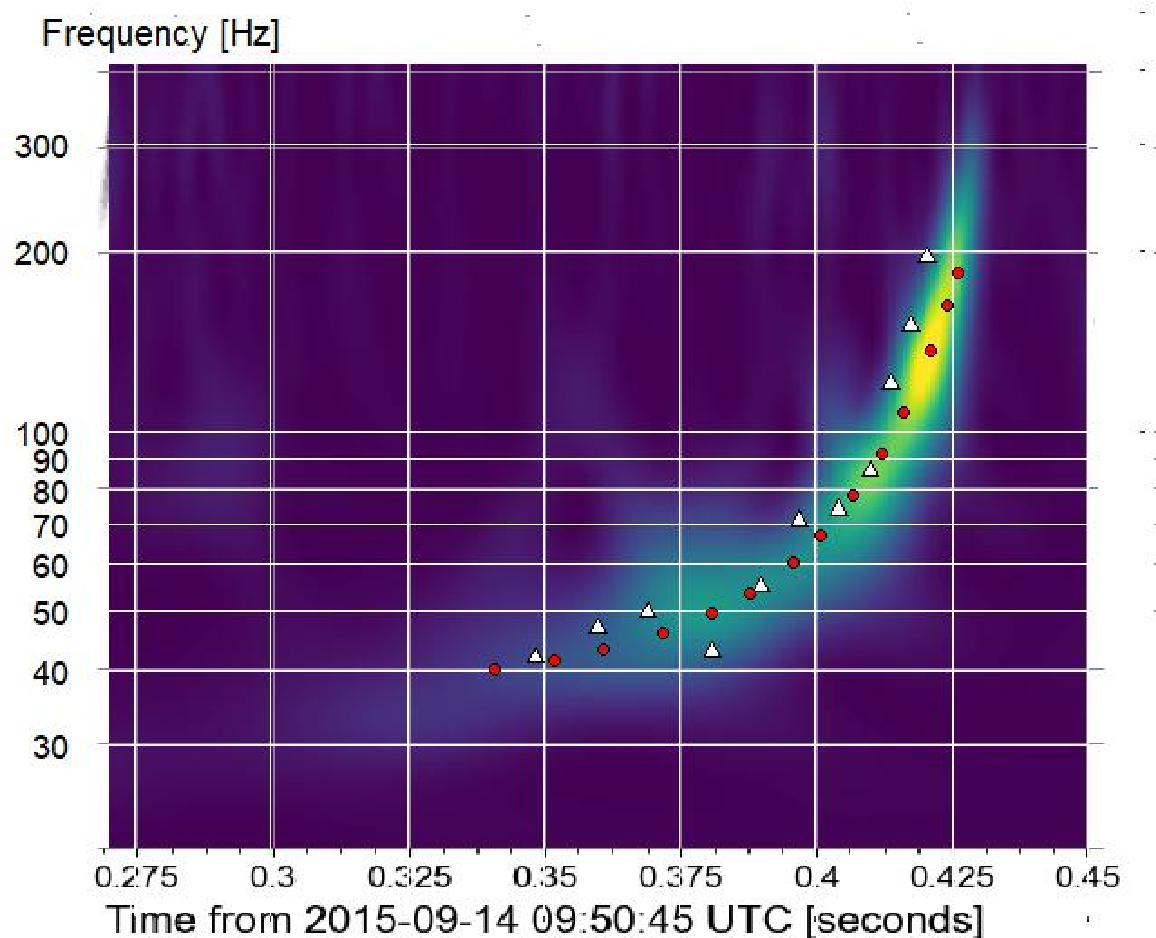}
\caption{{\it Left:} Instrumental strain of GW150914 from the LIGO 
Hanford H1 detector ($10^{-21}$ units);  
{\it Right:} time-frequency map for the H1 gravitational wave 
signal GW150914, with the white triangles indicating the frequencies at 
the successive zero-crossings of the instrumental strain time series used by the LVC 
for their calculations \cite{abbott17}; the red circles 
are the frequencies read directly from this diagram for our calculations.} 
\label{fig:time_freq_strain_map}
\end{figure}

After filtering the raw time-domain data recorded at a gravitational 
wave observatory, one can reveal instrumental strains representing 
gravitational wave signals produced by coalescing binary systems. 
The instrumental strain obtained by the LIGO H1
detector for GW150914 is shown on the left panel of  
\Fref{fig:time_freq_strain_map}. 
By matching of the LIGO H1 and L1  strains to NR templates, the LVC authors found that it was 
caused by an impact of a gravitational wave generated by two coalescing blackholes with 
masses $36^{+5}_{-4}$\,M$_\odot$ and $29\pm{4}$\,M$_\odot$, the source 
being at a luminosity distance $\sim 410$\,Mpc \cite{abbott16}.

The frequency growth in a GW signal can also be visualised 
in the form of a time-frequency map. Such  a map for the case 
of GW150914 is shown on the right panel of Figure\,\ref{fig:time_freq_strain_map}.  
The abscissa in both plots indicates time, the ordinate in the left plot 
indicates the instrumentally observed strain's amplitude (in units of $10^{-21}$). 
In the right plot, the normalised energy of the strain is colour-coded, with yellow 
corresponding to maximal normalised energies of the signal.
The abscissa in the right plot corresponds to the observed frequency. 

According to this time-frequency map, the frequency of the gravitational-wave 
signal  GW150914  increases from approximately 40\,Hz at $t \approx 0.33\,s$, when the 
signal becomes visible above the noise level, to approximately 180\,Hz at $t \approx 0.43\,s$, 
before the final coalescence of the objects. As with any quadrupole system, the GW frequency $f_{\mathrm{gw}}$ 
is the doubled  Keplerian orbital frequency of the system, $f_{\tt Kep}$, 
i.e. the maximal orbital frequency is supposed to be $f_{\tt Kep}\approx 85$\,Hz. 
The white triangles on the time-frequency map mark the GW frequencies corresponding to
the instrumental strain's zero-crossings, 
which were used by the LVC authors to estimate the characteristic (chirp) mass of GW150914 in their paper \cite{abbott17}. 

\noindent
The above plots have been produced by filtering the publicly available GW data maintained 
by the LVK collaboration (see the references in the {\it Data Availability} section).
But we have used a more convenient way of GW data manipulation: for this purpose there exists 
a specialised Python software package {\it GWpy} developed by D.M. Macleod, at al.
\cite{macleod21}. It provides the connectivity to the GW database, extraction of 
the requested fragment of row GW data and its appropriate filtering  and visualisation.

By integrating Eq.\,(\ref{freq2chirp_mass}), the LVC authors 
excluded the time-derivative $\dot{f}_\mathrm{gw}$:
\begin{equation}
f^{-8/3}_\mathrm{gw}(t) =  \frac{(8\pi)^{8/3}}{5}  \Big( \frac{G \mathcal{M}}{c^3}\Big)^{5/3} (t_c -t)
 \label{freq8_3}
\end{equation}
%
in order to fit linearly the quantities $f^{8/3}_\mathrm{gw}(t)$ corresponding to 
the zero-crossings of the instrumental strain, which gave them the chirp-mass estimate
$\mathcal{M}^{\tt LVC17} \approx 37$\,M$_\odot$ \cite{abbott17}. It differs from their more 
rigorous estimation $\mathcal{M}^{\tt LVC16} = 30.4^{+2.1}_{-1.9}$\,M$_\odot$
published earlier in \cite{abbott16}. The latter is based on the 
pre-calculated Numerical Relativity waveforms fitted to the whole instrumental strain. 
So, the low accuracy of $\mathcal{M}^{\tt LVC17}$ from \cite{abbott17} is not surprising, 
given that it is based on a small number of points sampled from 
a noisy signal.

\begin{figure}[htb]
\vspace{-0.2cm}
\hspace{2.5cm}
\includegraphics[scale=0.4]{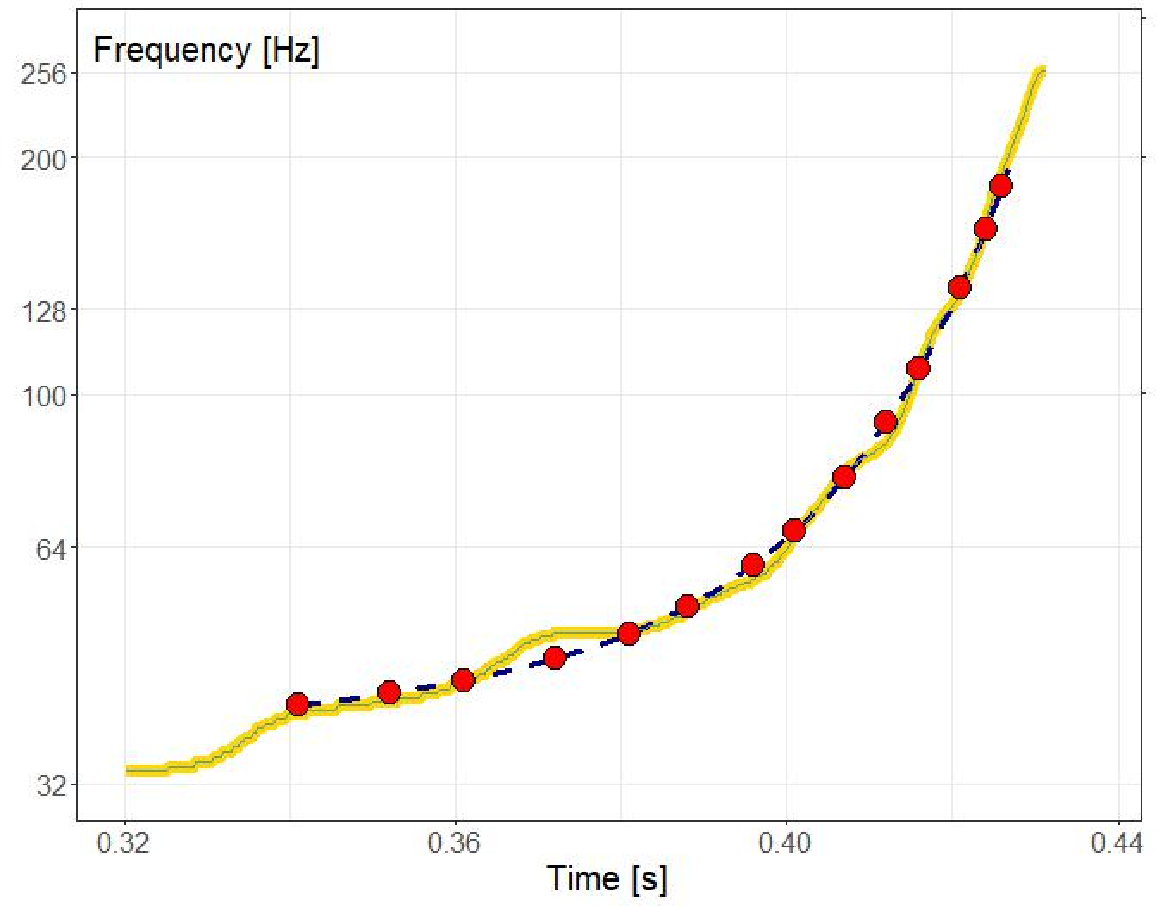}
\caption{Frequency growth of GW150914, as determined from the time-frequency 
map of the LIGO H1 signal. The yellow curve indicates the 
observed frequencies determined by the maximum of each frequency spectrum corresponding to
different moments of time.
These frequencies vary with time, and the smoothed frequency evolution 
(the dashed curve) gives a noise-reduced sequence of frequencies 
(red points) for calculating the chirp mass of the binary system. 
The smoothed frequency-evolution curve is obtained here by power-law 
fitting of the observed frequencies. The red points used for our calculations are also shown 
in the right plot of Figure 1.     
} 
\label{fig:time_freq_strain_fit}
\end{figure}

For more accurate determination of $\mathcal{M}$, 
we can read the frequencies $f_\mathrm{gw}$ directly from the 
time-frequency map of GW150914. 
This can be done, for example, by finding maxima  
of the spectral energy distributions at each time slot of the time-frequency map. 
We have extracted these spectral energy distributions and determined their maxima
by using the Python package {\it GWpy} \cite{macleod21}.
The frequencies corresponding to these maxima in the case of GW150914 are plotted 
as a yellow curve in \Fref{fig:time_freq_strain_fit}. 
 \begin{table}[ht]
 \caption{Chirp masses $\mathcal{M}_i$ estimated 
 for the moments of times taken from the time-frequency diagram for the
 instrumental strain of GW150914 
(red points on the top panel of \Fref{fig:time_freq_strain_map}): 
$t_i$ are the successive times;
 $f_\mathrm{gw}$ and $\dot{f}_\mathrm{gw}$
 are the observed GW frequencies and their time derivatives for each time interval;
 $r_i$ are the Keplerian distances between two objects at each time interval; 
 $f_{\tt Kep}$ are the corresponding Keplerian orbital frequencies 
 ($f_{\tt Kep}= 0.5\,f_\mathrm{gw}$); $v_i$  are the orbital velocities for
 the total mass of the system $M=67$\,M$_\odot$;  
 $\alpha_i^{\tt Sch}$ [\Eref{eq:timelapseSch}] and $\alpha_i^{\tt Kerr}$ [\Eref{eq:kerr_time_lapse}] are 
 the time-lapse factors for, respectively, the 
 Schwarzschild and Kerr metrics; $f^\mathrm{src}_\mathrm{gw}$ is the GW frequency translated to the source reference frame
 by using $\alpha^{\tt Kerr}$. 
}
\begin{adjustbox}{width=\columnwidth,center}
\medskip
 \label{cmass150914smoothed}
\begin{tabular}{@{}r|r|r|r|r|r|r|r|r|r|r|r}
\hline\noalign{\smallskip}
$i$~~~ & ~~~$t_i~~~~$   &   $~~~~f_\mathrm{gw}$~~       & $\dot{f}_\mathrm{gw}~~$  & ~~~$\mathcal{M}_i~~$ & $~~~~r_i~~~$  & ~$f_{\tt Kep}$~~ 
&  \multirow{2}{*}{~~~$v_i/c~$~} & \multirow{2}{*}{~~$\alpha_i^{\tt Sch}$~} & \multirow{2}{*}{~~$\alpha_i^{\tt Kerr}$~} 
& ~~~$f^{\tt src}_\mathrm{gw}$~ & ~~$\mathcal{M}^{\tt src}_i$~\\ 
 & $[s]~~~$ &   $[s^{-1}]~$ & ~~~$[s^{-2}]~~~$   & $~~~[M_\odot]~$  & [km]~~ & $[s^{-1}]~$ & & & & $[s^{-1}]$~~ & $[M_\odot]$~\\ 
\noalign{\smallskip}\hline
\tiny{1}~ & \tiny{2}~~~~ & \tiny{3}~~~~ & \tiny{4}~~~~~~ & \tiny{5}~~~~~ & \tiny{6}~~~~ 
& \tiny{7}~~~~ & \tiny{8}~~~ & \tiny{9}~~~~~ & \tiny{10}~~~ & \tiny{11}~~~~ & \tiny{12}~~ \\
\hline\noalign{\smallskip} 
{~1~} & ~~0.352~ & ~~~41.7~ & 134.9~   & 28.5~ & 811.2~ & 20.9~ & 0.177~ & 0.865~ & 0.806~ & 51.7~ &  --~~ \\
{~2~} & ~~0.361~ & ~~~43.3~ & 178.6~   & 31.1~ & 791.0~ & 21.7~ & 0.180~ & 0.861~ & 0.802~ & 54.0~ & 27.2~\\
{~3~} & ~~0.372~ & ~~~46.1~ & 256.4~   & 33.6~ & 758.4~ & 23.1~ & 0.183~ & 0.855~ & 0.793~ & 58.2~ & 29.1~\\
{~4~} & ~~0.381~ & ~~~49.7~ & 391.6~   & 36.8~ & 722.1~ & 24.8~ & 0.188~ & 0.847~ & 0.783~ & 63.4~ & 31.7~\\
{~5~} & ~~0.388~ & ~~~53.7~ & 577.4~   & 39.2~ & 685.4~ & 26.9~ & 0.193~ & 0.838~ & 0.771~ & 69.6~ & 33.4~\\
{~6~} & ~~0.396~ & ~~~60.7~ & 871.5~   & 38.3~ & 631.9~  & 30.3~ & 0.201~ & 0.823~ & 0.753~ & 80.6~ & 32.0~\\
{~7~} & ~~0.401~ & ~~~67.8~ & 1279.6~  & 38.7~ & 591.0~  & 33.5~ & 0.208~ & 0.809~ & 0.736~ & 91.2~ & 31.9~\\
{~8~} & ~~0.407~ & ~~~78.1~ & 1843.1~  & 34.4~ & 533.9~  & 39.1~ & 0.219~ & 0.786~ & 0.709~ & 110.2~ & 27.6~\\
{~9~} & ~~0.412~ & ~~~91.8~ & 2735.8~  & 30.6~ & 479.4~  & 45.9~ & 0.231~ & 0.758~ & 0.678~ & 135.4~ & 23.7~\\
{10~} & ~~0.416~ & ~~107.4~ & 3898.6~  & 26.8~ & 431.8~ & 53.7~ & 0.243~ & 0.726~ & 0.647~ & 166.0~ & 20.0~\\
{11~} & ~~0.421~ & ~~136.4~ & 5790.6~  & 20.1~ & 368.4~ & 68.2~ & 0.263~ & 0.668~ & 0.599~ & 227.7~ & 13.9~\\
{12~} & ~~0.424~ & ~~161.7~ & 8431.4~  & 17.3~ & 328.8~ & 80.8~ & 0.278~ & 0.616~ & 0.567~ & 285.0~ & 11.4~\\
{13~} & ~~0.426~ & ~~183.4~ & 10887.5~ & 15.3~ & 302.2~ & 91.7~ & 0.290~ & 0.570~ & 0.548~ & 334.9~ & 9.6~\\
\noalign{\smallskip}\hline\noalign{\smallskip}
\multicolumn{4}{r}{Average $\mathcal{M}~$:} & \multicolumn{2}{r}{30.1$\pm 2.4$\,$M_\odot$}~~ &  
\multicolumn{4}{r}{Average $\mathcal{M}^{\tt src}$:} & \multicolumn{2}{r}{24.3$\pm 2.6$\,$M_\odot$} \\
\noalign{\smallskip}\hline
\end{tabular}
\end{adjustbox} 
\end{table}

One can notice that the shape of this time-frequency curve 
is distorted with respect to its theoretical smoothness, which might be   
due to noise at the detector site. We can 
filter this noise by power-low fitting of the noisy curve, which  is shown 
in \Fref{fig:time_freq_strain_fit}) as the smoothed dashed curve. The filtered 
frequencies then can be read from this smoothed curve for any required moment of time.
Here we have taken 14 time-frequency points (red-coloured in Figures~\ref{fig:time_freq_strain_map} and 
\ref{fig:time_freq_strain_fit}) to have approximately the same number of 
observations as was used by by the LVC authors for their analysis.     
 
Then we can apply formula (\ref{freq2chirp_mass})
to each of these filtered frequencies, which are tabulated in the third column 
of Table~\ref{cmass150914smoothed}.  
This gives the average chirp mass $\mathcal{M} = 30.1\pm 2.4$\,M$_\odot$
(the bottom-left line of Table\,\ref{cmass150914smoothed})
determined with the use of the PN-approximation 
formula (\ref{freq2chirp_mass}), like in \cite{abbott17}. It coincides with 
the published value $\mathcal{M}^{\tt LVC16} = 30.4^{+2.1}_{-1.9}$\,M$_\odot$ \cite{abbott16}
obtained with the use of NR waveforms.

The sixth column of Table~\ref{cmass150914smoothed} 
gives the Keplerian distances $r_i$ between the components of this system 
at different moments of time. The peak amplitude of the GW signal
is achieved before the merger stage, at  $t_i= 0.42$\,s 
(the last row of Table~\ref{cmass150914smoothed}), with
the distance  $r_i$ between the objects being about 302~km.
Since the component masses of this system
are nearly equal \cite{abbott16,abbott17}, here we shall assume $m_1=m_2$ for simplicity.  
So, 
we get the individual blackhole masses 
$m_1=m_2= 2^{1/5} \mathcal{M} = 34.5$\,M$_\odot$
and the total mass $M =m_1+m_2 \approx 69$\,M$_\odot$.
The Keplerian orbital frequencies  
%
\begin{equation}
  f_{\tt Kep} = \frac{1}{2\pi}  \sqrt{\frac{GM}{r_{i}^3}} 
\label{eq_orb_freq}  
\end{equation} 
for each moment of time $t_i$ are given in seventh column 
of Table\,\ref{cmass150914smoothed}.
They correspond to the approximately Keplerian 
orbital motion of a mass $\mu = \frac{m_1 m_2}{m_1+m_2}=16.7$\,M$_\odot$
in the gravitational potential of a central 
mass $M = 67$\,M$_\odot$ in the equivalent one-body problem. 
For $t_i= 0.426$\,s,  the distance $r_i=302.2$\,km 
between two objects  is approximately equal to their three gravitational radii $r_g$.

\section{Equivalent one-body problem}

As we see from the sixth and eighth columns of Table\,\ref{cmass150914smoothed}, 
the Keplerian distances between the components of this system correspond to the relativistic regime 
of motion, $v_i/c \rightarrow 0.3$,  instead of  $v_i/c \ll 1$
required for the PN-regime  ($v_i$ being the orbital velocity 
given in column 8 of Table\,\ref{cmass150914smoothed} for each $i$-th moment of time).
Apparently, the condition (\ref{epsilon}) of the PN approximation is clearly not
met in the case of GW150914. 

Thence, we can have doubts with respect to the accuracy of the masses $\mathcal{M}_i$ 
computed with the use of the flat-spacetime formula (\ref{freq2chirp_mass}) 
because this formula, by design, does not account for the difference between the
coordinate reference frames in the vicinity of blackholes  (curved spacetime)
and near the detector (flat spacetime):  in both locations, 
 the PN-approximation assumes spacetime being flat.

Formula (\ref{freq2chirp_mass}) is based on the Einstein-Landau-Lifshitz quadru\-pole 
appro\-ximation (\ref{pn_energy_lvc2}) for which the quadrupole moments  
(\ref{pn_quad_lvc})--(\ref{pn_quad_lvc2}) and the components of the 
mass moment of inertia tensor (\ref{quad_inertia_moments}) are calculated 
by using the local (source) coordinates. 
In \cite{abbott17}, these coordinates are explicitly Newtonian, 
making no difference between the source and detector reference frames,
whilst, with distances between two objects being just a few  blackhole
gravitational radii, the deformation of the 
source reference frame  is substantial. 

In the classical one-body problem in curved spacetime, a mass 
$\mu=\frac{m_1 m_2}{m_1+m_2}$ moves in the Schwarzschild spacetime of the 
central mass $M=m_1+m_2$. In the test-mass limit, $\mu \rightarrow 0$, 
the orbiter does not deform the background Schwarzschild 
metric. This case has a theoretically {\it exact}
solution described in classical GR textbooks,
like \cite{zeldovich67,misner73,landau03}.
For example, among circular orbits around a Schwarzschild blackhole
there exists the innermost stable circular orbit (ISCO), whose radius
$r_{\tt ISCO} = 3r_g = 6GM/c^2 $ \cite{landau03,kaplan49}, which is 
a standard parameter characterising the Schwarzschild two-body problem.
The circular frequency of the test particle 
in this orbit is $\omega_{\tt \,ISCO} = 0.06804\, \frac{c^3}{GM}$ 
\cite{zeldovich67}.

For real cases, when the Schwarzschild metric is deformed by an orbiting mass 
$\mu > 0$, a theoretical solution (called the effective one-body formalism, EOB)
was found by A. Buonanno and T. Damour \cite{buonanno99} in the form of a post-Newtonian expansion with respect to 
the linearised Schwarzschild meric instead of flat spacetime.   
As was shown by these authors, the corrections to the Schwarzschild term 
in the EOB expansion are small. 
Even in the case of $r \simeq r_g$ and of the maximal value 
for the metric-deformation parameter $\nu$, the effective Schwarzschild radius 
is reduced to only $0.9277\, r_g$, which is not a significant change. 
In the case of the innermost stable circular orbit, 
the Schwarzschild term of the EOB PN-expansion is changed by 
only $2 \times 10^{-3}$.  As a result, for two equal blackhole masses
with the maximal metric-deformation parameter, the radius of the ISCO
appears to be only 5\% smaller, as compared to the standard Schwarzschild metric, i.e.,
$r_{\tt ISCO}^{\tt EOB} \simeq 5.7\,\frac{GM}{c^2}$
instead of  $6\,\frac{GM}{c^2}$. The circular frequency of the test particle 
in this orbit in the deformed Schwarzschild metric 
is $~\omega_{\tt \,ISCO}^{\tt \,EOB} = 0.07340\, \frac{c^3}{GM}$,
which is just a little bit higher than $0.06804\,\frac{c^3}{GM}$ 
in the standard Schwarzschild metric case. 

The above differences are not critical for our analysis, although they would need to be taken 
into account when developing an accurate algorithm for computing time-derivatives of frequencies. 
So, we shall continue 
with the standard GR-exact case of a circular orbit in the 
non-deformed Schwarzschild metric with its gravitational radius 
$r_g=2\,\frac{GM}{c^2}$.
In this case,  a circular orbit corresponds to the following values of the 
specific angular momentum 
$\tilde{\mathcal{L}} = \frac{p_\varphi}{\mu} = u_\varphi$:
\begin{equation}
  \tilde{\mathcal{L}} = \sqrt{\frac{rGM/c^2}{1 - 3GM/(rc^2)}}
\end{equation}
and specific energy ~~$\tilde{\mathcal{E}}= - \frac{p_t}{\mu} = -u_t$:
\begin{equation}
  \tilde{\mathcal{E}} = \frac{1- r_g/r}{\sqrt{1 - 3GM/(rc^2)}}\,.
\end{equation}
Consequently (see exercise 25.19 from \cite{misner73}), 
\begin{eqnarray} 
u^\varphi &=& \frac{d\varphi}{d\tau} = \frac{u_\varphi}{g_{\varphi \varphi}} = 
\frac{\tilde{\mathcal{L}}}{r^2}\,,\nonumber \\ 
u^t &=& \frac{dt}{d\tau} = \frac{u_t}{g_{tt}} = \frac{\tilde{\mathcal{E}}}{1- r_g/r},
\label{eq:dphidt}
\end{eqnarray}
where $g_{tt}$ and $g_{\varphi\varphi}$ 
are the spacetime metric coefficients
for, respectively,  time and the azimuthal angle in spherical coordinates, $\sqrt{g_{tt}}$ being 
the time lapse function.
Then, by calculating $d\varphi / dt$ 
from (\ref{eq:dphidt}), we get the angular frequency 
of orbital motion in the coordinate time $t$ (i.e., in the detector reference frame) 
\begin{equation}
 \omega^{\tt det} = \frac{d\varphi/d\tau}{dt/d\tau} = \frac{u^\varphi}{u^t} = \sqrt{\frac{GM}{c^2 r^3}},
 \label{eq:kepler_law}
\end{equation} 
which coincides with the Keplerian non-relativistic formula 
(that is, the lapse and shift functions are 
not involved), but which is an exact GR expression. 
This means that formula (\ref{eq:kepler_law}) fully accounts 
for the gravitational time dilatation effect 
in the vicinity of a mass $M$ located in the source reference
frame, though the laps function (the square root of the first
coefficient $g_{tt}$ in the spacetime metric tensor) is not involved.

It is for this reason that the analysis of the 
evolution of the system corresponding to the GW150914 signal  
led the LVC authors 
to the conclusion that, at any given moment of time, the motion of two coalescent masses 
of this system obeys Kepler laws until the very last orbits before the plunge phase.
And it is indeed the case, but only for the reference frame of a remote observer.

In accordance with formula (\ref{eq:kepler_law}), 
the orbital period $T^{\tt det} = 2\pi / \omega^{\tt det}$ reads
\begin{equation}
T^{\tt det} = \frac{2\pi\, c\, r^{3/2}}{\sqrt{GM}}\,.
\label{tkepler}
\end{equation}  
For example, by knowing $r_{\tt ISCO}= 6\,\frac{GM}{c^2}$ we can calculate 
the ISCO orbital period for a test mass $\mu$ as measured in the detector reference frame:
\begin{equation}
T_{\tt ISCO}^{\tt det} = \frac{12\pi}{\sqrt{2/3}} \frac{2GM}{c^3}\,.
\label{tisco}
\end{equation}  

\noindent
Seemingly, everything looks fine, as the formulae (\ref{freq2chirp_mass})  and
(\ref{eq:kepler_law}) give the source-frame (physical) mass $M=m_1+m_2$ of our binary blackhole
as derived from the detector-frame quantities 
$\omega^\mathrm{det}$, $f_\mathrm{gw}$ and $\dot{f}_\mathrm{gw}$,
whereas the motion of masses $m_1$ and $m_2$ is considered in the source reference frame. 
However, here we must make an important note that
the quadrupole moments and changes in the orbital-motion 
parameters  needed for obtaining $\dot{\omega}_\mathrm{orb}$ and the corresponding
$\dot{f}_\mathrm{gw}$ 
are calculated within the quadrupole-formalism framework, using local coordinates 
and proper time $\tau$ of the source. 

But sometimes, when calculating the quadrupole moments, this proper time
is denoted as $t$; i.e. the distinction between $\tau$ and $t$ is dropped.   
This is explicit in the LVC PN-analysis \cite{abbott17} when, for example,  
formulae (\ref{pn_quad_lvc}) to (\ref{pn_energy_lvc2}) are used for 
the derivation of formula (\ref{freq2chirp_mass}). In that derivation, 
the authors, indeed, do not make any distinction between the proper and
coordinate time. 

It would seem that it is of no much importance, since each orbit is Keplerian, whilst 
the Keplerian formula (\ref{eq:kepler_law})  is exact in the GR framework
and gives the orbital frequency in the detector reference frame when using
the physical mass $M$ of the system. That is, the knowledge of the 
quantities depending on the local time in the source reference frame 
seems to be unnecessary. 

However, it is unnecessary only at a first glance, because, when we calculate 
the orbit evolution due to the loss of energy (\ref{pn_energy_lvc2})
carried away by GWs, then the change of the orbital radius $r$ and the 
corresponding change in $\dot{\omega}_\mathrm{orb}$ of the orbital
frequency  $\omega_\mathrm{orb}$ must be calculated with the use 
of the physical time of the system -- and not with the coordinate time of 
a remote observer. 

That is why without knowing the time derivative of the orbital frequency 
 $\dot{\omega}_\mathrm{orb}^\mathrm{src}$ in the source reference frame,
 but using instead the observed orbital frequency 
 $\omega_\mathrm{orb}^\mathrm{det} = \pi f_\mathrm{gw}$
and the observed time derivative  $\dot{f}_\mathrm{gw}$ (both related to the coordinate time $t$),
one will obtain from  (\ref{freq2chirp_mass}) a non-physical estimate
of mass $\mathcal{M}$ for the detector reference frame, 
whereas formula (\ref{eq:kepler_law}) does not work at all
without knowing $\dot{\omega}_\mathrm{orb}$.

The situation with the absence of the lapse and shift functions in 
formula (\ref{eq:kepler_law}) is similar to the situation with the absence of these 
functions in the Arnowitt-Deser-Misner (ADM) parametrisation \cite{arnowitt59,arnowitt60},
which is used, in particular, for rigorous consideration of motion 
of compact binary systems within the PN-approximation of GR
\cite{schafer85,schafer18}, as well as for using the PN-approximation in the EOB
formalism  for pre-calculating accurate GW waveforms when the
motion of coalescing binary blackhole systems approach to the relativistic      
regime and the distance between the components of the system diminishes 
to one gravitational radius \cite{buonanno99}. 
 
In the ADM+PN scheme, the energy-momentum tensor is integrated
over the entire space. Then the reduced action functional is varied,
which gives the equations of motion with no lapse and 
shift functions involved -- similar to the exact GR-formula (\ref{eq:kepler_law}).
The radiation-reaction force, as well as the PN-expansion coefficients, are 
calculated in the ADM scheme afterwards. Thus, since in these latter calculations 
the lapse and shift functions are not involved, the 
waveforms obtained within the ADN+EOB formalism are calculated 
for the reference frame of a remote observer, like in the case of the 
exact GR-formula (\ref{eq:kepler_law}).  
 
This non-involvement of the time-lapse  function in the ADM equations of motion is usually 
interpreted as the capacity of the ADM formalism to fully 
account for any gravitationally 
induced time-dilatation effect by computing two PN-expansions: 
one for the near wave zone, and another -- for the far zone
(in the asymptotically-flat spacetime). Both 
PN-expansions are then matched, which is understood  as the accounting for the 
gravitational time dilatation near the source. However, as we have already 
found for the exact GR-formula case (\ref{eq:kepler_law}), since the lapse function
is not involved when calculating both PN-expansions, they do not allow 
obtaining the physical value of the binary system's  
chirp-mass without using the physical time in the source reference frame.

Here we shall continue with the theoretically exact case 
of the GR Keplerian formula  (\ref{eq:kepler_law}) for the Schwarzschild metric
because it gives a more clear picture of the indispensability of
using the physical $\tau$ instead of the coordinate time $t$ 
when computing the time derivative of the orbital frequency in order 
to avoid a systematic error in $\mathcal{M}$. In fact, here we do not need
exact waveforms from the ADM+EOB formalism for highlighting the 
issue with the mentioned systematic error.    
%

%
Time and its conjugate variable ener\-gy (aka freq\-uency) in the distorted reference frame 
near a blackhole are known to be fundamentally 
different from the time and energy in the flat spacetime 
of a remote observer. 
This, for instance, was mentioned by Ya.B. Zeldovich and I.D. Novikov 
in their textbook \cite{zeldovich67}, in the paragraph discussing 
ISCO orbits. They wrote, that, since all the processes 
in a gravitational field  are slowing down for a remote observer  by the factor 
$\alpha=\sqrt{g_{tt}}$, then the ISCO frequency in the 
detector reference frame will be smaller than in the source reference frame
and will correspond to formula (\ref{tisco}), which is taken from the mentioned
Zeldovich and Novikov's textbook. 
Here $\alpha$ is the time-lapse function which gives the factor 
$\sqrt{2/3}$ in the denominator of formula (\ref{tisco}) for the ISCO case.  

The information about the energy-loss due to the GW emission and about 
the corresponding evolution of a binary blackhole system with mass $M$ 
is contained in the observed parameter $\dot{f}_\mathrm{gw}$. 
Therefore, in order to properly estimate the parameter $M$ we  
need to translate $\dot{f}_\mathrm{gw}$ 
from the detector reference frame to the source reference frame.
The derivative of the frequency, translated to the source reference frame 
will correspond to the physical time and physical 
parameters of motion of the system.   
In the first approximation,  
this can be done by applying the aforementioned time-lapse function.
We shall describe this below in the next two Sections.

\section{Time-lapse function}

\noindent
By using the value of mass $M=69$\,M$_\odot$, which we have currently accepted for
the system under consideration, and applying formula (\ref{tisco}), we can find that    
the frequency $f_{\tt ISCO}=1/T_{\tt ISCO}^{\tt det}$ for this system must be equal 
to $f^{\tt det}_{\tt ISCO}=31.8$\,Hz in the detector reference frame. 
If we convert the ISCO orbital period from the detector to source reference frame
by multiplying (\ref{tisco}) by the factor $\sqrt{2/3}$:
\begin{equation}
T_{\tt ISCO}^{\tt src} = \frac{12\pi \cdot 2GM}{c^3}\,,
\label{tiscosrc}
\end{equation}
we get $f^{\tt src}_{\tt ISCO}=38.9$\,Hz, which exceeds
$f^{\tt det}_{\tt ISCO}$ by about 22\%. This converted 
frequency exceeds also the (detector-frame) frequency $f^{\tt EOB}_{\tt ISCO}= 35.40$\,Hz 
as estimated by using the correction for the Schwarzschild metric deformation  
given by \cite{buonanno99} for the maximal value of the metric-deformation parameter 
due to a non-zero value of the test mass $\mu$.  
Thus, when passing from the detector to source reference frame, one gets 
orbital frequencies, which are higher than those corrected for the 
Schwarzschild metric deformation in the EOB formalism.    

Accordingly, the frequencies $f_\mathrm{gw}$ and their derivatives $\dot{f}_\mathrm{gw}$ from Table\,\ref{cmass150914smoothed} 
should be increased if we want to convert them into the physical frequencies in the 
source reference frame,
$f_\mathrm{gw} \rightarrow f^\mathrm{src}_\mathrm{gw}$ and 
$\dot{f}_\mathrm{gw} \rightarrow \dot{f}^\mathrm{src}_\mathrm{gw}$.
Therefore, the chirp-mass $\mathcal{M}^{\tt src}$ estimated for GW150914 
via formula (\ref{freq2chirp_mass}) by using the physical values $f^\mathrm{src}_\mathrm{gw}$ 
and $\dot{f}^\mathrm{src}_\mathrm{gw}$ instead of the observed ones will be smaller than
our initial estimation of $\mathcal{M}$
shown on the bottom-left line of Table\,\ref{cmass150914smoothed}. 
It will also be smaller than the values of $\mathcal{M}$ 
published by \cite{abbott17} because, as we see, those 
estimates were also made for the detector reference frame.  

In principle, the LVC scheme for processing GW signals envisages the conversion 
of the mass $\mathcal{M}^{\tt det}$ from the detector reference frame to the 
mass  $\mathcal{M}^{\tt src}$ in the source reference frame, but this conversion 
is made only with the purpose of accounting for the cosmological redshift, i.e., for the time dilatation
proportional to the distance to the observed object (due to expansion of the Universe).
However, as we can see from our analysis,  this conversion to the source reference frame 
 is incomplete.      

If we use the GW150914 signal as it is, in its form at the detector output,  
then we (the remote observer) see Keplerian orbital motion of two bodies
with masses $\sim$34.5\,M$_\odot$ and with the adiabatically decreasing orbit radius
and orbital period, as if these two bodies were moving in flat spacetime. 
This, of course, corresponds to the conditions of the PN approximation, but  
evidently does not conform  to real physical conditions of curved spacetime near 
the components of GW150914.

At the final stages of their orbital motion, two coalescing objects are located very
closely to each other, e.g. when their gravitational radii are $\sim 100$\,km, the radius 
of their common circular orbit is only $\sim 150$\,km. That is, the coordinate reference 
frame in the vicinity of this orbit is essentially distorted by the energy-mass of the
objects.  
Even at the initial moment of time $t_i=0.35\,s$ (the first row of Table\,\ref{cmass150914smoothed}),
the distance between two objects (811\,km) is only 1.3 times larger than the 
ISCO radius in the equivalent one-body problem with  
$M=m_1+m_2 \approx 69$\,M$_\odot$. This means that relativistic corrections 
related to source reference frame are essential for the whole period of the 
observation of the  GW150914 signal.     

In this particular case, by assuming the total mass of the binary system to be $M=69$\,M$_\odot$,
as derived from the chirp-mass $\mathcal{M}=30.1$\,M$_\odot$,
we get the ISCO radius $r_{\tt ISCO} \simeq 612$\,km, which corresponds to a value $r_i$
between the sixth and seventh rows of Table\,\ref{cmass150914smoothed}.
Concurrently, in units of proper time, the orbital period will be $T_{\tt ISCO} \approx 0.0314\,s$,
corresponding to the orbital frequency $f^\mathrm{src}_\mathrm{orb} \approx 39$\,Hz  and to the 
GW frequency $f^\mathrm{src}_\mathrm{gw} = 2 f^\mathrm{src}_\mathrm{orb}= 78$\,Hz.    
By contrast, in our Table\,\ref{cmass150914smoothed} the frequency $f$ for the orbit with radius 612\,km is between 
60.7\,Hz and 67.8\,Hz (sixth and seventh rows of the Table), which is much lower than 78\,Hz. 
For the moment of time $t_i=0.426\,s$ 
in the last row of Table\,\ref{cmass150914smoothed}, 
and for the initial parameters $M=69$\,M$_\odot$, $r_g \approx 204$\,km and $r_i=302$\,km,
the Schwarzschild time-lapse function is 
\begin{equation}
\alpha^{\tt Sch} = \sqrt{1- \frac{r_g}{r}} = \sqrt{1- \frac{204}{302}}=0.57\,
\label{eq:timelapseSch}
\end{equation}
which is quite significant.
This is yet another indication that the estimated mass
$M=69$\,M$_\odot$ derived from the tabulated frequencies $f_\mathrm{gw}$ does not conform 
to the proper physical time of GW150914; that is, our initial value of  $\mathcal{M}$ 
was overestimated. 

Accordingly, in order to properly find the unbiased characteristic mass $\mathcal{M}^\mathrm{src}$
of GW150914, the frequency $f_\mathrm{gw}=183$\,Hz for $i=13$
must be translated to the source reference frame by dividing it 
by the factor (\ref{eq:timelapseSch}). 
This gives $f^\mathrm{src}_\mathrm{gw} \approx 321$\,Hz instead of 
$f^\mathrm{det}_\mathrm{gw} =183$\,Hz.

\section{Angular momentum}

In the previous paragraph, we have estimated the correction for 
the time dilatation in the static Schwarzschild metric. 
However, a static metric cannot realistically describe a {\it rotating}
two-body system. Therefore, we need to estimate the time-lapse function
 $\alpha$ using the more appropriate Kerr metric, in which the 
orbital period, as expressed via the proper time, is smaller in comparison with the coordinate time 
not only due to the time dilatation, but also due to the effect 
of frame-dragging under rotation: see, e.g., \S33.4 from the textbook
by \cite{misner73}, volume 3, as well as \S4.3 from \cite{frolov98} 
or  \S104 from \cite{landau03}.  
As our considerations here are approximate (we are presenting a proof 
of existence of a systematic error), we shall continue regarding the mapping 
of our two-body system to an equivalent one-body system with a Kerr-like metric. 
(we assume that the components of the binary system are tidally locked).
With this,    
\begin{equation} \label{eq:freq_src}
 \omega_\mathrm{src} = \frac{\omega_\mathrm{det}}{\sqrt{g_{tt}+2  \omega_\mathrm{det} g_{t\varphi} 
 +\omega_\mathrm{det}^2 g_{\varphi \varphi}}}\,  
\end{equation}
for the proper time of an observer moving in the source reference frame
Here $\omega_\mathrm{src}=2\pi f_{\tt orb}^\mathrm{src}$ is the orbital 
angular frequency in the 
source reference frame, and the orbital angular frequency
for a remote observer is $\omega_\mathrm{det}=\pi f_{\tt det}$
[formula (\ref{eq:freq_src}) is written for the 
equatorial plane, $\theta=\frac{\pi}{2}$].
The time-lapse function 
\begin{equation}
\alpha^{\tt Kerr} =  \sqrt{g_{tt}+2 \omega_\mathrm{det} g_{t\varphi} +\omega_\mathrm{det}^2 g_{\varphi \varphi}}\, 
\label{eq:kerr_time_lapse}
\end{equation}
-- the denominator of formula (\ref{eq:freq_src}) -- corresponds to the Kerr metric \cite{landau03}:
\begin{eqnarray} 
ds^2 = & \left(1 - \frac{r_g}{r}\right) c^2 dt^2 
  - \left(r^2 + a^2 + \frac{r_ga^2}{r} \right) d\varphi^2 +  \\
  &+  \frac{2r_g a}{r} d\varphi dt
  - \frac{r^2}{\Delta} dr^2 - r^2 d\theta^2\,,
\label{eq:kerr_metrics}
\end{eqnarray}
written in the covariant coordinates for $\theta=\frac{\pi}{2}$ and with the notations
$\Delta = r^2 - r_g r + \frac{a^2}{c^2}$
and $a = J /Mc$, where $a$ is the specific angular momentum of 
an isolated Kerr blackhole regarded in the equivalent one-body problem.

The Kerr metric describes the geometry of spacetime around a rotating black hole. 
In the case of a binary blackhole system, the geometry of spacetime can be described by 
a Kerr-like metric, which is a perturbation of the Kerr metric
due to the presence of a second blackhole. The Kerr metric can be written in Boyer-Lindquist 
coordinates as:
\begin{eqnarray} 
ds^2 = & -\left(1 - \frac{2GM}{r\rho^2}\right) c^2 dt^2 - \frac{4GJ}{rc^2}\sin^2 \theta dt d\varphi 
        + \frac{\rho^2}{\Delta} dr^2 + \rho^2 d\theta^2 \nonumber \\
       & + \left(r^2 +a^2 +\frac{2GJ^2}{rc^2\rho^2}\right) \sin^2\theta d\varphi^2\,,\nonumber
\label{eq:kerr_metricBoyer}
\end{eqnarray}
where $M$ is the mass of the black hole, J is its angular momentum, 
$a = J/Mc$ is its spin parameter, $G$ is the gravitational constant, $c$ is the speed of light, 
$t$ is the time coordinate, $r$ is the radial coordinate, $\vartheta$ is the polar angle,
$\varphi$ is the azimuthal angle, and
\begin{eqnarray} 
\rho^2 = & r^2 +a^2 \cos \theta, \nonumber \\
 \Delta =  & + r^2 - 2Gr +a^2\,. \nonumber
\label{eq:kerr_rhoDelta}
\end{eqnarray}
In the case of a binary black hole system, we can write the Kerr-like metric as:
\begin{eqnarray} 
ds^2 = & - \left(1- \frac{2Gm_1}{r_1\rho_1^2} - \frac{2Gm_2}{r_2\rho_2^2}\right) c^2 dt^2
         - \frac{4GJ_1}{r_1c^2 \rho_1^2}\sin^2 \theta_1 dt d\varphi \nonumber \\
       &  - \frac{4GJ_2}{r_2c^2\rho_2^2}\sin^2 \theta_2 dt d\varphi_2  
         + \frac{\rho_1^2}{\Delta_1} dr_1^2 +   \frac{\rho_2^2}{\Delta_2} dr_2^2 
         +\rho_1^2 d\theta_1^2 + \rho_2^2 d\theta_2^2 \nonumber \\ 
        & +\left(r_1^2 +a_2^2 + \frac{2GJ^2_1}{r_1 c^2 \rho_1^2}\right) \sin^2 \theta_1 d\varphi_1^2 
        + \left( r_2^2 + a_2^2 + \frac{2GJ_2^2}{r_2 c^2\rho_2^2} \right)\sin^2\theta_2 d\varphi_2^2 \nonumber \\
       & - \frac{4GJ_{12}}{r_{12} c^2 \rho_1 \rho_2}\sin \theta_1 \sin \theta_2 d\varphi_1\varphi_2\,, \nonumber
\label{eq:kerr_rhoDelta}
\end{eqnarray}
where $m_1$, $m_2$, $J_1$, $J_2$, $a_1$, $a_2$, $r_1$, $r_2$, $\theta_1$, $\theta_2$, 
$\varphi_1$, $\varphi_2$, $\varphi_3$, $\rho_1$, $\rho_2$, $\Delta_1$, $\Delta_2$,
$r_{12}$ and $J_{12}$ are the mass, angular momentum, spin
parameter, radial coordinate, polar angle, azimuthal angle, distance 
from the first black hole, distance from the second
black hole, the radial distance between the two black holes, and the angular momentum due to their interaction,
respectively.
The terms involving $m_1$ and $m_2$ represent the gravitational field of each black hole, while the term
involving $J_{12}$ describes the interaction between them. The interaction term only appears 
in the cross term $d\varphi_1$ $d \varphi_2$.
which couples the azimuthal angles of the two black holes. The radial distances $r_1$ and $r_2$ 
are related to the distance from the center of mass of the system $r$ and the mass ratio 
$q = m_2/m_1$ as follows:
\begin{equation*}
 r_1 = \frac{r}{1+q} \mathrm{~~~~~~and~~~~~~} r_2=\frac{qr}{1+q}. 
\label{eq:r1r2}
\end{equation*}
The distance between two black holes is given by:
\begin{eqnarray}
r_{12}  = & [ (r_1 \sin \theta_1 \cos \varphi_1 - r_2 \sin \theta_2 \cos \varphi)^2 \nonumber\\  
   &+(r_1 \sin \theta_1 \sin \varphi_1 - r_2 \sin \theta_2 \sin \varphi_2 )^2 \nonumber \\
    &+  (r_1 \cos \theta_1 -r_2 \cos \theta_2 )^2 ]^{1/2} \nonumber
\label{eq:r12}
\end{eqnarray}%
and the interaction term $J_{12}$ is given by:
\begin{equation*}
 J_{12} = \frac{m_1 m_2}{m_1 + m_2} \sqrt{\frac{G q (1+q)r}{1+q^2} } 
\label{eq:intrc_term}
\end{equation*}
The Kerr-like metric for a binary blackhole system is a highly non-linear 
and complicated equation that cannot be solved analytically in general. 
However, numerical methods can be used to study the dynamics of the system and
predict the gravitational waves emitted during the inspiral motion. Thus, 
the angular momentum $a$ for binary system can be expressed, as in the Kerr metric 
for an isolated black hole, the specific angular momentum $a$ being given 
$a= J/Mc$, where $J$ is the angular momentum of the blackhole, $M$ is its mass, 
and $c$ is the speed of light.
Passing back from the equivalent one-body problem to the two-body problem, 
the specific angular momentum $a$ of each black hole is given by
$a_1 = J_1/(m_1c)$ and $a_2 = J_2/(m_2c)$, where $J_1$ and $J_2$ are the angular 
momenta of twoblackholes, and $m_1$ and $m_2$ are their masses. Note that in
the binary black hole system, each black hole has its own specific angular 
momentum $a$, which is different from the total angular momentum $J$ of the system. 
The total angular momentum of the binary black hole system is given by
\begin{equation*}
J= J_1 + J_2 + J_{12}\,,
\label{eq:totangmom}
\end{equation*}
where $J_{12}$ is the angular momentum due to the orbital motion of the two black 
holes around each other, which has been expressed above.

For our simplified calculations with the tidally locked blackholes, 
we compute the values $\alpha^{\tt Kerr}$ via formula 
(\ref{eq:kerr_time_lapse}) with the use of the metric components  
(\ref{eq:kerr_metrics}) $g_{tt} =  1- r_g/r$,~ 
$g_{\varphi \varphi} =  -(r^2 + a^2  + a^2r_g/r)$~
and $g_{t\varphi} =   2a r_g / (rc^2)$
are given in the 10-th column of Table\,\ref{cmass150914smoothed}.
As one would expect, these values are smaller than the values 
$\alpha^{\tt Sch}$ for the static Schwarzschild metric (9-th column of the 
same Table). The frequencies $f^{\tt src}_\mathrm{gw}$ of the GW signal in the source reference frame,
as well as the chirp-masses $\mathcal{M}_i^{\tt src}$ computed with the use of the 
corrected values $f^{\tt src}_\mathrm{gw}$ and $\dot{f}^{\tt src}_\mathrm{gw}$ 
are given in columns 11 and 12  of Table\,\ref{cmass150914smoothed}.  

When computing the values $\alpha^{\tt Kerr}$, we have taken into account
the mass-energy corresponding to the angular momentum $J$ of the two-body system that produced the 
GW150914 signal.
This energy makes its own contribution to the spacetime curvature. We have also taken into account  
the gradual loss of the mass-energy dissipated by gravitational waves.
For this purpose, we have used the expression for the emitted gravitational-wave 
energy from a circular orbit  \cite{zeldovich67}:
\begin{equation}
\frac{dE}{dt} = 0.2 \frac{c^5}{G}\left( \frac{\mu}{M} \right)^2 \left(1+\frac{\mu}{M}\right)
\left(\frac{r_g}{r}\right)^5\,.
\label{eq:gw_mass_loss}
\end{equation}   
By taking (\ref{eq:gw_mass_loss}) into account, we find that the mass-loss of this system   
due to the GW emission amounts to approximately 3.8\,M$_\odot$ when the system arrives at its last 
evolutionary stage. Accordingly, as the orbital parameters of the system
evolve, the specific angular momentum $a$ and the gravitational radius $r_g$ used for
the calculation of the metric components (\ref{eq:kerr_metrics}) gradually decrease,
which we have taken into account in our calculations as well.  

As the frequencies of the physical orbital motion are higher than the observed ones,
the chirp-masses in the source reference frame given in the 12-th column of 
Table\,\ref{cmass150914smoothed}, as well as their mean value, 
$\mathcal{M}^{\tt src}=$24.3\,M$_\odot$, are substantially smaller 
than our initial estimate $\mathcal{M}^{\tt det}=$30.1\,M$_\odot$ made 
without accounting for the slowdown of physical processes in a gravitational field.
Accordingly, the corrected individual masses of the blackholes in the system     
GW150914 are also reduced from  34.5\,M$_\odot$, which we saw in \S3, to approximately 27.9\,M$_\odot$
(they will be further reduced when applying the cosmological redshift
correction, but we shall limit our considerations here to only the gravitational 
redshift correction).

\section{Discussion}
\subsection{Other cases from GWTC}
Our analysis of the two-body problem based on the example of the coalescing binary blackhole system 
that produced the GW150914 signal leads us to the conclusion that the estimation of orbital parameters 
and masses of these blackholes requires a correction for the time dilatation, causing a 
slowdown of all processes in the gravitational field of the system as seen from the 
detector reference frame. 

\begin{figure}[ht]
\vspace{-0.1cm}
\hspace{1cm}
\includegraphics[scale=1.0]{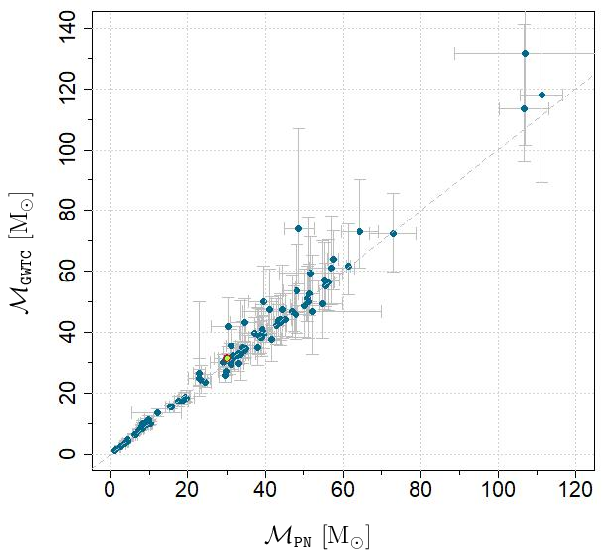}
\caption{Diagonal plot showing the correspondence between the 97 
chirp masses $\mathcal{M}_{\tt GWTC}$ (detector-frame)  from the Gravitational-Wave Transient Catalogue 
(vertical axis) 
and those calculated by using the PN-flat-spacetime formula (\ref{freq2chirp_mass}) -- the horizontal axis.  
The yellow point indicates the chirp mass value corresponding to the GW150914 signal discussed here.    
} 
\label{fig:pn_nr_correspondence}
\end{figure}

According to our calculations, this correction in the case of GW150914  
is quite large -- it exceeds the 
correction for the cosmological redshift, and also the correction for the 
Schwarzschild metric deformation within the framework of the EOB formalism.
If one  assumes the frequencies of the GW signal being the same in the source and detector 
reference frames, then the calculation of the chirp-masses $\mathcal{M}$ of the so-far discovered 
coalescent binary blackhole systems will result in these masses being overestimated (we disregard for the moment 
the cosmological redshift, which is usually accounted for at the end of the GW signal 
processing, when the chirp-masses are already calculated). 

The most telling proof that this is indeed the case can be obtained by taking the 
characteristic (chirp) masses of coalescing binaries 
published by the LIGO\,/\,Virgo collaboration 
\cite{abbott19,venumadhav20, nitz20,abbott20a, abbott20b, abbott20c}
and by comparing them with the chirp masses computed for the same systems but with the 
use of the post-Newtonial approximation formula (\ref{freq2chirp_mass}),
which is inherently, by design, a flat-spacetime formula. 
Such a comparison is presented in a diagonal plot of
Figure\,\ref{fig:pn_nr_correspondence}, where the published detector-frame chirp masses 
go along the vertical axis, and the chirp masses computed 
by using the flat-spacetime formula (\ref{freq2chirp_mass}) are along the horizontal 
axis.   
The Pearson correlation test (performed by using the function {\it cor.test(x, y)} 
from the R-software package {\it stats}) gives the following result:
\begin{itemize}
\item the correlation coefficient is: $0.985^{+0,004}_{-0.008}$
\item 
p-value $< 2.2 \cdot 10^{-16},$
\end{itemize}
which indicates the practically perfect diagonality of the 
plot in Fig.\,\ref{fig:pn_nr_correspondence} and 
demonstrates the full one-to-one correspondence between the masses $\mathcal{M}$
computed by two different methods. This means that the GW waveforms calculated
within the ADM+EOB approach  (corresponding to the masses along the 
vertical axis of Figure\,\ref{fig:pn_nr_correspondence})
lack the gravitationally induced time-lapse correction because the flat-spacetime 
formula (\ref{freq2chirp_mass}) applied to the detector-frame frequencies for calculating
the detector-frame chirp masses lacks this correction by design.  
 
\subsection{Distance scaling}
 The detected amplitudes of GW signals produced by coalescing binary systems 
 are used for estimating distances to these systems by comparing the theoretical 
 GW amplitude with the observed amplitude.  
 The theoretical amplitude of a GW signal is proportional to the $\frac{5}{3}$-rd power 
 of the binary system's chirp mass (Equation \ref{freq2chirp_mass}). Thus, there is a 
theoretical dependence between the estimated chirp masses and luminosity distances to the 
observed binary systems. Thence, if chirp masses are overestimated due to a systematic
error, then the luminosity distances will be overestimated as well.   

\begin{figure}[ht]
\vspace{-1.0cm}
\hspace{2.5cm}
\includegraphics[scale=0.7]{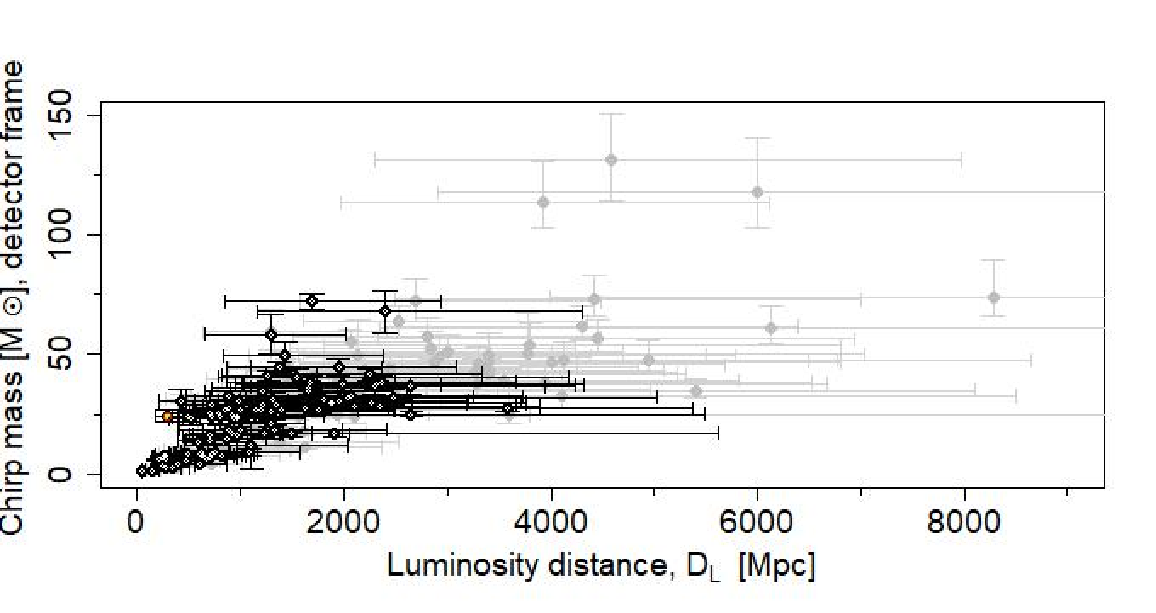}
\vspace{-0.3cm}
\caption{Chirp-masses $\mathcal{M}_{\tt corr}$ (vertical axis) 
of coalescing binary black hole systems detected during the  LVC/LVK O1, O2  and O3 observing runs
corrected for the gravitational redshift (black points), and the correspondingly  
corrected luminosity distances $D_L^{\tt corr}$ of these systems (horizontal axis) .
The grey points indicate the original chirp masses and luminosity distances from GWTC,
for comparison purposes. The yellow point corresponds to the corrected chirp mass and luminosity 
distance of the GW150914 source.} 
\label{fig:gw_mass_dist2}
\end{figure}

We can rescale these overestimated distances by using the ratio between the 
corrected and uncorrected chirp masses: 
\begin{equation}
  D_L^c = \left( \frac{\mathcal{M}^c}{\mathcal{M}} \right)^{5/3} D_L, 
  \label{dist_scaling}
\end{equation}
where $D_L^c$ is the scaled GW luminosity distance, $D_L$ is the published distance, 
$\mathcal{M}$ and $\mathcal{M}^c$ are the original and corrected chirp masses of a 
coalescing binary system.  
The rescaled distances $D_L^c$ to the 97 GW sources presented in Table~\ref{tab_apptable1} 
are shown in column 8 of this Table, together with the corrected chirp masses $\mathcal{M}^c$
shown in column 7 (we have used the rescaling factor $(\mathcal{M}^c/\mathcal{M})^{5/3}$
from formula (\ref{dist_scaling}).
The corrected relationship between $D_L^c$ and $\mathcal{M}^c$ is plotted in 
Figure~\ref{fig:gw_mass_dist2} with black points. The original (uncorrected) chirp masses 
of GWTC sources and their corresponding luminosity distances are indicated by the grey points
in this plot.   

As one would expect, the corrected luminosity distances 
are now much smaller, most of them being closer than $\sim 3000$\,Mpc, whereas the uncorrected 
luminosity distances are twice as large. 
Still one can notice some systematic relationship between the corrected 
luminosity distances and chirp masses of the GWTC sources (in the case of the uncorrected 
chirp masses and luminosity distances, such a systematic relationship is plainly seen).  

Although one would expect a certain observational selection effect, the remaining 
linear correlation between these two parameters seems to indicate that our
lapse-function corrections that we applied here for chirp mass corrections might be too approximate. 
But we did not set our goal for devising the rigorous corrections. At this moment, it suffice 
to present a proof of existence of such an error and our approximate estimations accomplish this 
task. Meanwhile, more accurate determination of this error would depend on deeper analysis in the future.

\section{Conclusion} 
Here we have presented a proof of existence of a systematic error in the estimation of 
the characteristic (chirp) masses of coalescing binary systems discovered via gravitational 
wave detections. An important implication of this result is our conclusion that 
the estimated luminosity distances to the discovered gravitational sources are largely overestimated,
especially in the cases with large chirp masses. 

Since the locations on the sky of gravitational sources are currently determined with 
large uncertainties, the search for any possible electromagnetic transient counterparts
is performed within the locations of host-galaxy candidates pre-selected on the basis 
of information about the likely gravitational source distances. Therefore, more accurate 
knowledge of this distances is crucial for increasing the probability of finding 
electromagnetic counterparts of gravitational wave sources. So far, only one such 
counterpart was found.

\section*{Acknowledgements}
This research has made use of data or software obtained from the Gravitational 
Wave Open Science Center (gw-openscience.org), a service of LIGO Laboratory, 
the LIGO Scientific Collaboration, the Virgo Collaboration, and KAGRA. LIGO Laboratory 
and Advanced LIGO are funded by the United States National Science Foundation (NSF) 
as well as the Science and Technology Facilities Council (STFC) of the United Kingdom, 
the Max-Planck-Society (MPS), and the State of Niedersachsen/Germany for support of the 
construction of Advanced LIGO and construction and operation of the GEO600 detector. 
Additional support for Advanced LIGO was provided by the Australian Research Council. 
Virgo is funded, through the European Gravitational Observatory (EGO), by the French Centre 
National de Recherche Scientifique (CNRS), the Italian Istituto Nazionale di Fisica Nucleare 
(INFN) and the Dutch Nikhef, with contributions by institutions from Belgium, Germany, Greece, 
Hungary, Ireland, Japan, Monaco, Poland, Portugal, Spain. The construction and operation of KAGRA 
are funded by Ministry of Education, Culture, Sports, Science and Technology (MEXT), and Japan 
Society for the Promotion of Science (JSPS), National Research Foundation (NRF) and Ministry 
of Science and ICT (MSIT) in Korea, Academia Sinica (AS) and the Ministry of Science and 
Technology (MoST) in Taiwan. In particular we have made use of the following archives: 

\noindent
GWTC-1: A Gravitational-Wave Transient Catalog of Compact Binary Mergers 
Observed by LIGO and Virgo during the First and Second Observing Runs 
\cite{abbott19};  
GWTC-2.1: Deep Extended Catalog of Compact Binary Coalescences Observed by
LIGO and Virgo During the First Half of the Third Observing Run
\cite{abbott21a};
GWTC-3:  Compact Binary Coalescences Observed by LIGO and Virgo During the
Second Part of the Third Observing Run 
\cite{abbott21,abbott21c};
2-OGC: Open Gravitational-wave Catalog of binary mergers from analysis of public 
Advanced LIGO and Virgo data \cite{nitz20} and
the list of blackholes found in Galactic X-ray binaries \cite{wiktorowicz13}.

\noindent
We also acknowledge the use for our work of the Python software research package {\it GWpy} developed by 
D.M. Macleod, J.S. Areeda, S.B. Coughlin, T.J. Massinger and A.L. Urban \cite{macleod21}. 
This package is available at \url{https://gwpy.github.io/docs/stable/}. 

\noindent
We are thankful for the contribution to this paper of three anonymous
reviewers whose numerous suggestions led to the essential improvement in the revised version of our manuscript.
We especially acknowledge the help provided by the first reviewer who prepared 
a text for the improvement of \S6 discussing the angular momentum.    

\noindent  
We would like to thank Dr. Leslie Morrison, Dr. Alice Breeveld, Dr. N.P.M. Kuin, 
Prof. Mat Page, and Prof. Sergei Soloviev for useful discussions
on the matters in this paper.

\appendix
\section{Chirp mass comparison}

Table~\ref{tab_apptable1} compares the chirp masses $\mathcal{M}_\mathrm{GWTC}$ of the gravitational wave 
sources published in the GWTC catalogues with  the chirp masses $\mathcal{M}_\mathrm{PN}$ of the same sources, 
as calculated here by using the quadrupole post-Newtonian formula (\ref{freq2chirp_mass}).
This comparison is also graphically presented in Figure\,\ref{fig:pn_nr_correspondence}. 
The comparison is also made between the published source luminosity distances, $D_L^\mathrm{GWTC}$ and   
the luminosity distances $D_L^\mathrm{c}$ corrected by taking into account the systematic error in the 
chirp-mass estimation, see \Eref{dist_scaling}. 

\setcounter{table}{0}
\begin{table}[hb]
\vspace{-0.5cm}
 \caption{Chirp masses of the gravitational wave sources from the published GWTC catalogues, 
 $\mathcal{M}_\mathrm{GWTC}$, in comparison with the chirp masses $\mathcal{M}_\mathrm{PN}$ 
 of the same sources as  calculated here by using the quadrupole post-Newtonian formalism. 
 The masses are given for the detector reference frame and are expressed in M$_\odot$. 
 Column 7 gives the post-Newtonian chirp masses corrected by using of the time lapse function 
 $\alpha^{\mathrm{Kerr}}$ (\ref{eq:kerr_time_lapse}).  Columns 4 and 8 give, respectively, the published 
 source luminosity distances, $D_L^\mathrm{GWTC}$ and the corrected ones, $D_L^\mathrm{c}$ 
 (both are in Gpc).
}
\medskip
\begin{adjustbox}{width=\columnwidth,center}
\medskip
 \label{tab_apptable1}
\begin{tabular}{ccrrcccc}
\hline
N   &  ID  &  $\mathcal{M}_\mathrm{GWTC}$ & $D_L^\mathrm{GWTC}$ & {\scriptsize Reference} & $\mathcal{M}_\mathrm{PN}$ & $\mathcal{M}_\mathrm{PN}^\mathrm{c}$ &  $D_L^\mathrm{c}$ \\
\noalign{\smallskip}\hline
\hline
\tiny{1}~ & \tiny{2}~~~~ & \tiny{3}~~~~ & \tiny{4}~~~~~~ & \tiny{5}~~ & \tiny{6}~~~~ & \tiny{7}~~ & \tiny{8}~~\\
\noalign{\smallskip}\hline
\noalign{\smallskip}
01  &  GW150914\_095045 &  $31.17 ^{+3.0 }_{   -4.4}$ &  $0.44 ^{+0.15 }_{-0.17 }   $ &  \cite{abbott19} & $ 30.07  \pm{ 2.35 }$ &  $24.28\pm{2.57  }$ & 0.29$^{+0.10}_{-0.11}$     \\
02  &  GW151012\_095443 &  $18.39 ^{+2.1 }_{   -1.2}$ &  $1.08 ^{+0.55 }_{-0.49 }   $ &  \cite{abbott19} & $ 19.61  \pm{ 2.03 }$ &  $16.66\pm{1.88  }$ & 0.92$^{+0.47}_{-0.42}$     \\
03  &  GW151226\_033853 &  $9.70  ^{+0.3 }_{   -0.3}$ &  $0.45 ^{+0.18 }_{-0.19 }   $ &  \cite{abbott19} & $  8.93  \pm{ 1.29 }$ &  $7.29 \pm{0.93  }$ & 0.28$^{+0.11}_{-0.12}$     \\
04  &  GW170104\_101158 &  $25.68 ^{+2.2 }_{   -1.8}$ &  $0.99 ^{+0.44 }_{-0.43 }   $ &  \cite{abbott19} & $ 29.61  \pm{ 1.52 }$ &  $24.41\pm{1.65  }$ & 0.91$^{+0.40}_{-0.39}$     \\
05  &  GW170121\_212536 &  $31.37 ^{+4.0 }_{   -3.0}$ &  $1.24 ^{+0.87 }_{-0.71 }   $ & \cite{venumadhav20} & $ 31.49  \pm{ 4.64 }$ &  $24.00\pm{3.64  }$ & 0.79$^{+0.55}_{-0.46}$     \\
06  &  GW170304\_163753 &  $48.87 ^{+8.0 }_{   -7.0}$ &  $2.91 ^{+1.47 }_{-1.32 }   $ & \cite{venumadhav20} & $ 50.06  \pm{ 3.80 }$ &  $34.74\pm{4.32  }$ & 1.65$^{+0.83}_{-0.74}$     \\
07  &  GW170608\_020116 &  $8.45  ^{+0.2 }_{   -0.2}$ &  $0.32 ^{+0.12 }_{-0.11 }   $ &  \cite{abbott19} & $ 8.05   \pm{ 0.80 }$ &  $7.56 \pm{0.87  }$ & 0.26$^{+0.10}_{-0.09}$     \\
08  &  GW170727\_010430 &  $44.18 ^{+6.0 }_{   -6.0}$ &  $2.44 ^{+1.27 }_{-1.08 }   $ & \cite{venumadhav20} & $ 43.66  \pm{ 2.96 }$ &  $28.99\pm{2.96  }$ & 1,21$^{+0.63}_{-0.53}$     \\
09  &  GW170729\_185629 &  $52.75 ^{+6.5 }_{   -4.8}$ &  $2.84 ^{+1.40 }_{-1.36 }   $ &  \cite{abbott19} & $ 51.29  \pm{ 2.74 }$ &  $38.38\pm{2.69  }$ & 1.68$^{+0.82}_{-0.80}$     \\
10  &  GW170809\_082821 &  $29.88 ^{2.1+ }_{   -1.7}$ &  $1.03 ^{+0.32 }_{-0.39 }   $ &  \cite{abbott19} & $ 28.95  \pm{ 3.17 }$ &  $23.94\pm{3.08  }$ & 0.71$^{+0.22}_{-0.27}$     \\
11  &  GW170814\_103043 &  $26.99 ^{1.4+ }_{   -1.1}$ &  $0.60 ^{+0.15 }_{-0.22 }   $ &  \cite{abbott19} & $ 29.79  \pm{ 2.00 }$ &  $23.68\pm{1.91  }$ & 0.48$^{+0.12}_{-0.18}$     \\
12  &  GW170817\_124104 &  $1.198 ^{+0.001}_{ -0.001}$ & $0.040^{+0.007}_{-0.015}   $ &  \cite{abbott19} & $  1.19  \pm{ 0.011}$ &  $ 1.18\pm{0.011 }$ & 0.042$^{+0.003}_{-0.007}$  \\
\hline
\end{tabular}
\end{adjustbox}
\end{table}

\clearpage
\setcounter{table}{0}
\begin{table}
 \caption{~(continuation)}
\medskip 
\begin{adjustbox}{width=\columnwidth,center}
\medskip
\begin{tabular}{ccrrcccc}
  \hline
N   &  ID  &  $\mathcal{M}_\mathrm{GWTC}$ & $D_L^\mathrm{GWTC}$ & {\scriptsize Reference} & $\mathcal{M}_\mathrm{PN}$ & $\mathcal{M}_\mathrm{PN}^\mathrm{c}$ &  $D_L^\mathrm{c}$ \\
\noalign{\smallskip}\hline
\hline
\tiny{1}~ & \tiny{2}~~~~ & \tiny{3}~~~~ & \tiny{4}~~~~~~ & \tiny{5}~~ & \tiny{6}~~~~ & \tiny{7}~~ & \tiny{8}~~ \\
\noalign{\smallskip}\hline
\noalign{\smallskip}
13  &  GW170818\_022509 &  $32.06 ^{+2.1  }_{  -1.7}$ &  $1.06 ^{+0.42 }_{-0.38 }   $ &  \cite{abbott19} & $ 32.94  \pm{ 1.25 }$ &  $25.17\pm{1.18  }$ & 0.71$^{+0.28}_{-0.25}$     \\
14  &  GW170823\_131358 &  $39.42 ^{+4.6  }_{  -3.6}$ &  $1.94 ^{+0.97 }_{-0.90 }   $ &  \cite{abbott19} & $ 39.57  \pm{ 2.13 }$ &  $31.18\pm{2.49  }$ & 1.31$^{+0.66}_{-0.61}$     \\
15  &  GW190403\_051519 &  $74.12 ^{+15.1 }_{  -8.4}$ &  $8.28 ^{+6.72 }_{-4.29 }   $ & \cite{abbott21a} & $ 48.68  \pm{ 3.90 }$ &  $32.16\pm{4.37  }$ & 2.06$^{+1.67}_{-1.07}$     \\
16  &  GW190408\_181802 &  $24.60 ^{+1.9  }_{  -1.2}$ &  $1.55 ^{+0.40 }_{-0.60 }   $ &  \cite{abbott20} & $ 24.58  \pm{ 1.32 }$ &  $21.19\pm{1.25  }$ & 1.30$^{+0.33}_{-0.50}$     \\
17  &  GW190412\_053044 &  $15.30 ^{+0.4  }_{  -0.3}$ &  $0.74 ^{+0.14 }_{-0.17 }   $ &  \cite{abbott20} & $ 15.28  \pm{ 1.18 }$ &  $13.25\pm{1.11  }$ & 0.58$^{+0.11}_{-0.13}$     \\
18  &  GW190413\_052954 &  $39.11 ^{+5.5  }_{  -4.1}$ &  $3.55 ^{+2.27 }_{-1.66 }   $ &  \cite{abbott20} & $ 38.29  \pm{ 1.03 }$ &  $29.89\pm{0.82  }$ & 2.27$^{+1.45}_{-1.06}$     \\
19  &  GW190413\_134308 &  $56.43 ^{+8.2  }_{  -5.4}$ &  $4.45 ^{+2.48 }_{-2.12 }   $ &  \cite{abbott20} & $ 56.19  \pm{ 1.59 }$ &  $38.50\pm{2.46  }$ & 2.36$^{+1.31}_{-1.12}$     \\
20  &  GW190421\_213856 &  $46.49 ^{+5.9  }_{  -4.2}$ &  $2.88 ^{+1.37 }_{-1.38 }   $ &  \cite{abbott20} & $ 47.28  \pm{ 2.89 }$ &  $37.27\pm{3.12  }$ & 1.99$^{+0.95}_{-0.95}$     \\
21  &  GW190424\_180648 &  $43.09 ^{ +5.8 }_{ -4.6  }$ &      $2.20 ^{+1.58}_{-1.16}   $ &  \cite{abbott20} & $34.76 \pm{3.80 }$ &  $26.91\pm{4.01 }$ & 1.00$^{+0.72}_{-0.53}$ \\
22  &  GW190425\_081805 &  $1.48  ^{ +0.02}_{ -0.02 }$ &      $0.16 ^{+0.07}_{-0.07}   $ &  \cite{abbott20} & $1.44  \pm{0.003}$ &  $1.42 \pm{0.002}$ & 0.14$^{+0.06}_{-0.06}$ \\
23  &  GW190426\_152155 &  $2.60  ^{ +0.08}_{ -0.08 }$ &      $0.37 ^{+0.18}_{-0.16}   $ &  \cite{abbott20} & $2.55  \pm{0.11 }$ &  $2.45 \pm{0.14 }$ & 0.34$^{+0.16}_{-0.14}$ \\
24  &  GW190426\_190642 &  $31.48 ^{ +19.1}_{ -17.4 }$ &      $4.58 ^{+3.40}_{-2.28}   $ & \cite{abbott21a} & $106.95\pm{18.3 }$ &  $72.23\pm{3.08 }$ & 1.69$^{+1.25}_{-0.84}$ \\
25  &  GW190503\_185404 &  $38.35 ^{ +4.2 }_{ -4.2  }$ &      $1.45 ^{+0.69}_{-0.63}   $ &  \cite{abbott20} & $38.73 \pm{1.61 }$ &  $30.46\pm{1.86 }$ & 0.98$^{+0.47}_{-0.43}$ \\
26  &  GW190512\_180714 &  $18.54 ^{ +1.2 }_{ -1.0  }$ &      $1.43 ^{+0.55}_{-0.55}   $ &  \cite{abbott20} & $19.23 \pm{1.37 }$ &  $16.82\pm{1.09 }$ & 1.22$^{+0.47}_{-0.47}$ \\
27  &  GW190513\_205428 &  $29.59 ^{ +3.8 }_{ -1.9  }$ &      $2.06 ^{+0.88}_{-0.80}   $ &  \cite{abbott20} & $33.24 \pm{1.07 }$ &  $26.86\pm{1.22 }$ & 1.75$^{+0.75}_{-0.68}$ \\
28  &  GW190514\_065416 &  $47.60 ^{ +7.9 }_{ -4.8  }$ &      $4.13 ^{+2.65}_{-2.17}   $ &  \cite{abbott20} & $41.01 \pm{3.58 }$ &  $30.32\pm{3.42 }$ & 1.94$^{+1.25}_{-1.02}$ \\
29  &  GW190517\_055101 &  $49.74 ^{ +4.0 }_{ -4.0  }$ &      $1.86 ^{+1.62}_{-0.84}   $ &  \cite{abbott20} & $31.09 \pm{1.35 }$ &  $24.03\pm{1.56 }$ & 0.97$^{+0.84}_{-0.44}$ \\
30  &  GW190519\_153544 &  $64.08 ^{ +6.4 }_{ -7.1  }$ &      $2.53 ^{+1.83}_{-0.92}   $ &  \cite{abbott20} & $57.47 \pm{1.38 }$ &  $44.40\pm{2.17 }$ & 1.37$^{+0.99}_{-0.50}$ \\
31  &  GW190521\_030229 & 1$05.09 ^{ +17.0}_{ -10.6 }$ &      $3.92 ^{+2.19}_{-1.95}   $ &  \cite{abbott20} & $106.7 \pm{6.3  }$ &  $58.45\pm{8.1  }$ & 1.30$^{+0.72}_{-0.64}$ \\
32  &  GW190521\_074359 &  $39.80 ^{ +3.2 }_{ -2.5  }$ &      $1.24 ^{+0.40}_{-0.57}   $ &  \cite{abbott20} & $39.59 \pm{1.67 }$ &  $32.31\pm{2.15 }$ & 0.88$^{+0.28}_{-0.40}$ \\
33  &  GW190527\_092055 &  $34.99 ^{ +9.1 }_{ -4.2  }$ &      $2.49 ^{+2.48}_{-1.24}   $ &  \cite{abbott20} & $37.97 \pm{1.39 }$ &  $29.64\pm{1.63 }$ & 1.89$^{+1.88}_{-0.94}$ \\
34  &  GW190602\_175927 &  $72.17 ^{ +9.1 }_{ -8.5  }$ &      $2.69 ^{+1.79}_{-1.12}   $ &  \cite{abbott20} & $72.86 \pm{6.06 }$ &  $49.36\pm{5.86 }$ & 1.42$^{+0.95}_{-0.59}$ \\
35  &  GW190620\_030421 &  $57.06 ^{ +8.3 }_{ -6.5  }$ &      $2.81 ^{+1.68}_{-1.31}   $ &  \cite{abbott20} & $55.23 \pm{3.68 }$ &  $34.51\pm{3.05 }$ & 1.21$^{+0.72}_{-0.57}$ \\
36  &  GW190630\_185205 &  $29.38 ^{ +2.1 }_{ -2.1  }$ &      $0.89 ^{+0.56}_{-0.37}   $ &  \cite{abbott20} & $31.03 \pm{1.08 }$ &  $25.34\pm{1.34 }$ & 0.69$^{+0.44}_{-0.29}$ \\
37  &  GW190701\_203306 &  $55.21 ^{ +5.4 }_{ -4.9  }$ &      $2.06 ^{+0.76}_{-0.73}   $ &  \cite{abbott20} & $55.39 \pm{1.69 }$ &  $41.25\pm{2.26 }$ & 1.27$^{+0.47}_{-0.45}$ \\
38  &  GW190706\_222641 &  $73.02 ^{+10.0 }_{ -7.0  }$ &      $4.42 ^{+2.59}_{-1.93}   $ &  \cite{abbott20} & $64.40 \pm{4.85 }$ &  $44.61\pm{3.36 }$ & 1.94$^{+1.14}_{-0.85}$ \\
39  &  GW190707\_093326 &  $9.86  ^{ +0.6 }_{ -0.5  }$ &      $0.77 ^{+0.38}_{-0.37}   $ &  \cite{abbott20} & $9.42  \pm{0.42 }$ &  $8.48 \pm{0.40 }$ & 0.59$^{+0.29}_{-0.28}$ \\
40  &  GW190708\_232457 &  $15.57 ^{ +0.9 }_{ -0.6  }$ &      $0.88 ^{+0.33}_{-0.39}   $ &  \cite{abbott20} & $15.86 \pm{0.79 }$ &  $13.64\pm{0.79 }$ & 0.70$^{+0.26}_{-0.31}$ \\
41  &  GW190719\_215514 &  $38.54 ^{ +6.5 }_{ -4.0  }$ &      $3.94 ^{+2.59}_{-2.00}   $ &  \cite{abbott20} & $39.03 \pm{3.27 }$ &  $28.20\pm{2.29 }$ & 2.34$^{+1.54}_{-1.19}$ \\
42  &  GW190720\_000836 &  $10.32 ^{ +0.5 }_{ -0.8  }$ &      $0.79 ^{+0.69}_{-0.32}   $ & \cite{abbott20} & $10.04 \pm{0.46 }$ &  $8.90 \pm{0.53 }$ & 0.62$^{+0.54}_{-0.25}$ \\
43  &  GW190725\_174728 &  $8.88  ^{ +0.5 }_{ -0.5  }$ &      $1.03 ^{+0.52}_{-0.43}   $ & \cite{abbott20} & $8.10  \pm{0.38 }$ &  $7.03 \pm{0.59 }$ & 0.70$^{+0.35}_{-0.29}$ \\
44  &  GW190727\_060333 &  $44.33 ^{ +5.3 }_{ -3.7  }$ &      $3.30 ^{+1.54}_{-1.50}   $ & \cite{abbott20} & $45.00 \pm{2.94 }$ &  $30.87\pm{3.10 }$ & 1.81$^{+0.84}_{-0.82}$ \\
45  &  GW190728\_064510 &  $10.15 ^{ +0.5 }_{ -0.3  }$ &      $0.87 ^{+0.26}_{-0.37}   $ & \cite{abbott20} & $10.29 \pm{0.49 }$ &  $9.21 \pm{0.49 }$ & 0.75$^{+0.22}_{-0.32}$ \\
46  &  GW190731\_140936 &  $45.72 ^{ +7.1 }_{ -5.2  }$ &      $3.30 ^{+2.39}_{-1.72}   $ & \cite{abbott20} & $47.73 \pm{2.82 }$ &  $36.69\pm{3.23 }$ & 2.29$^{+1.66}_{-1.19}$ \\
47  &  GW190803\_022701 &  $42.32 ^{ +5.7 }_{ -4.1  }$ &      $3.27 ^{+1.95}_{-1.58}   $ &  \cite{abbott20} & $42.76 \pm{2.54 }$ &  $31.86\pm{2.63 }$ & 2.04$^{+1.21}_{-0.98}$ \\
48  &  GW190805\_211137 &  $61.25 ^{ +8.8 }_{ -6.3  }$ &      $6.13 ^{+3.72}_{-2.38}   $ & \cite{abbott21a} & $57.06 \pm{5.74 }$ &  $36.89\pm{3.52 }$ & 2.64$^{+1.60}_{-1.62}$ \\
49  &  GW190814\_211039 &  $6.39  ^{+0.06 }_{ -0.06 }$ &      $0.24 ^{+0.04}_{-0.05}   $ & \cite{abbott20} & $6.20  \pm{0.28 }$ &  $5.85 \pm{0.26 }$ & 0.21$^{+0.04}_{-0.04}$ \\
50  &  GW190828\_063405 &  $34.5  ^{ +3.4 }_{ -2.1  }$ &      $2.13 ^{+0.66}_{-0.93}   $ & \cite{abbott20} & $34.89 \pm{2.25 }$ &  $26.74\pm{2.64 }$ & 1.39$^{+0.43}_{-0.61}$ \\
51  &  GW190828\_065509 &  $17.29 ^{ +1.2 }_{ -1.0  }$ &      $1.60 ^{+0.62}_{-0.60}   $ & \cite{abbott20} & $18.75 \pm{1.37 }$ &  $16.53\pm{1.20 }$ & 1.48$^{+0.57}_{-0.56}$ \\
52  &  GW190909\_114149 &  $50.06 ^{+17.2 }_{ -7.5  }$ &      $3.77 ^{+3.27}_{-2.22}   $ & \cite{abbott20} & $51.25 \pm{1.03 }$ &  $37.34\pm{1.87 }$ & 2.31$^{+2.00}_{-1.36}$ \\
53  &  GW190910\_112807 &  $43.90 ^{ +4.1 }_{ -4.1  }$ &      $1.46 ^{+1.03}_{-0.58}   $ &  \cite{abbott20} & $43.27 \pm{1.16 }$ &  $28.71\pm{1.30 }$ & 0.72$^{+0.51}_{-0.28}$ \\
\hline
\end{tabular}
\end{adjustbox}
\end{table}

\clearpage
\setcounter{table}{0}
\begin{table}
 \caption{~(continuation)}
 \medskip
\begin{adjustbox}{width=\columnwidth,center}
\medskip
\begin{tabular}{ccrrcccc}
  \hline
N   &  ID  &  $\mathcal{M}_\mathrm{GWTC}$ & $D_L^\mathrm{GWTC}$ & {\scriptsize Reference} & $\mathcal{M}_\mathrm{PN}$ & $\mathcal{M}_\mathrm{PN}^\mathrm{c}$ &  $D_L^\mathrm{c}$ \\
\noalign{\smallskip}\hline
\hline
\tiny{1}~ & \tiny{2}~~~~ & \tiny{3}~~~~ & \tiny{4}~~~~~~ & \tiny{5}~~ & \tiny{6}~~~~ & \tiny{7}~~ & \tiny{8}~~ \\
\noalign{\smallskip}\hline
\noalign{\smallskip}
54  &  GW190915\_235702 &  $32.89 ^{ +3.2 }_{ -2.7  }$ &      $1.62 ^{+0.71}_{-0.61}   $ &  \cite{abbott20} & $33.18 \pm{4.14 }$ &  $23.50\pm{4.24 }$ & 0.92$^{+0.40}_{-0.35}$ \\
55  &  GW190916\_200658 &  $47.61 ^{ +8.2 }_{ -5.4  }$ &      $4.94 ^{+3.71}_{-2.38}   $ & \cite{abbott21a} & $44.29 \pm{2.23 }$ &  $28.42\pm{2.49 }$ & 2.09$^{+1.57}_{-1.01}$ \\
56  &  GW190917\_114630 &  $4.26  ^{ +0.2 }_{  -0.2 }$ &      $0.72 ^{+0.30}_{-0.31}   $ & \cite{abbott21a} & $4.16  \pm{0.20 }$ &  $3.85 \pm{0.18 }$ & 0.61$^{+0.25}_{-0.26}$ \\
57  &  GW190924\_021846 &  $6.50  ^{ +0.2 }_{  -0.2 }$ &      $0.57 ^{+0.22}_{-0.22}   $ &  \cite{abbott20} & $6.41  \pm{0.57 }$ &  $5.77 \pm{0.59 }$ & 0.47$^{+0.18}_{-0.18}$ \\
58  &  GW190925\_232845 &  $18.56 ^{ +1.1 }_{  -1.1 }$ &      $0.93 ^{+0.46}_{-0.35}   $ & \cite{abbott21a} & $19.33 \pm{1.01 }$ &  $15.76\pm{0.56 }$ & 0.70$^{+0.35}_{-0.26}$ \\
59  &  GW190926\_050336 &  $37.82 ^{ +9.0 }_{  -4.9 }$ &      $3.28 ^{-3.40}_{-1.73}   $ & \cite{abbott21a} & $41.43 \pm{2.58 }$ &  $31.90\pm{3.27 }$ & 2.47$^{+2.56}_{-1.30}$ \\
60  &  GW190929\_012149 &  $49.40 ^{ +14.9}_{  -8.2 }$ &      $2.13 ^{+3.65}_{-1.05}   $ & \cite{abbott20} & $54.70 \pm{5.08 }$ &  $40.62\pm{3.54 }$ & 1.54$^{+2.63}_{-0.76}$ \\
57  &  GW190924\_021846 &  $6.50  ^{+0.2 }_{-0.2 }$ &   $0.57^{+0.22}_{-0.22}   $ & \cite{abbott20} & $6.41  \pm{0.57 }$ &  $5.77 \pm{0.59 }$ & 0.47$^{+0.18}_{-0.18}$ \\
58  &  GW190925\_232845 &  $18.56 ^{+1.1 }_{-1.1 }$ &   $0.93^{+0.46}_{-0.35}   $ & \cite{abbott21a} & $19.33 \pm{1.01 }$ &  $15.76\pm{0.56 }$ & 0.70$^{+0.35}_{-0.26}$ \\
59  &  GW190926\_050336 &  $37.82 ^{+9.0 }_{-4.9 }$ &   $3.28^{-3.40}_{-1.73}   $ & \cite{abbott21a} & $41.43 \pm{2.58 }$ &  $31.90\pm{3.27 }$ & 2.47$^{+2.56}_{-1.30}$ \\
60  &  GW190929\_012149 &  $49.40 ^{+14.9}_{-8.2 }$ &   $2.13^{+3.65}_{-1.05}   $ & \cite{abbott20} & $54.70 \pm{5.08 }$ &  $40.62\pm{3.54 }$ & 1.54$^{+2.63}_{-0.76}$ \\
61  &  GW190930\_133541 &  $9.77 ^{+0.5 }_{-0.5  }$ &   $0.76^{+0.36}_{-0.32}   $ & \cite{abbott20} & $9.76  \pm{0.68 }$ &  $8.57 \pm{0.68}$ & 0.61$^{+0.29}_{-0.26}$ \\
62  &  GW191103\_012549 &  $10.01^{+0.66}_{-0.57 }$ &   $0.99^{+0.50}_{-0.47}   $ & \cite{abbott21} & $8.22  \pm{0.54 }$ &  $7.53 \pm{0.52}$ & 0.62$^{+0.31}_{-0.29}$ \\
63  &  GW191105\_143521 &  $9.62 ^{+0.61}_{-0.45 }$ &   $1.15^{+0.43}_{-0.48}   $ & \cite{abbott21}& $8.25  \pm{1.12 }$ &  $7.87 \pm{1.00}$ & 0.82$^{+0.31}_{-0.34}$ \\
64  &  GW191109\_010717 &  $59.38^{+9.6 }_{-7.5  }$ &   $1.29^{+1.13}_{-0.65}   $ & \cite{abbott21}& $51.66 \pm{8.04 }$ &  $30.34\pm{4.70}$ & 0.42$^{+0.37}_{-0.21}$ \\
65  &  GW191113\_071753 &  $13.48^{+1/1 }_{-1.0  }$ &   $1.37^{+1.15}_{-0.62}   $ & \cite{abbott21} & $12.03 \pm{6.50 }$ &  $11.85\pm{9.73}$ & 1.10$^{+0.92}_{-0.49}$ \\
66  &  GW191126\_115259 &  $11.24^{+0.95}_{-0.71 }$ &   $1.62^{+0.74}_{-0.74}   $ & \cite{abbott21} & $9.84  \pm{1.10 }$ &  $8.80 \pm{1.29}$ & 1.08$^{+0.50}_{-0.50}$ \\
67  &  GW191127\_050227 &  $46.94^{+11.7}_{-9.1  }$ &   $3.4 ^{+3.1 }_{-1.9 }   $ & \cite{abbott21} & $52.09 \pm{17.94}$ &  $26.26\pm{11.2}$ & 1.29$^{+1.18}_{-0.72}$ \\
68  &  GW191129\_134029 &  $8.48 ^{+0.43}_{-0.28 }$ &   $0.79^{+0.26}_{-0.33}   $ & \cite{abbott21} & $8.23  \pm{0.19 }$ &  $7.45 \pm{0.14}$ & 0.64$^{+0.21}_{-0.26}$ \\
69  &  GW191204\_110529 &  $26.53^{+3.6 }_{-3.3  }$ &   $1.8 ^{+1.7 }_{-1.1 }   $ & \cite{abbott21} & $23.04 \pm{1.55 }$ &  $18.51\pm{1.43}$ & 0.99$^{+0.99}_{-0.60}$ \\
70  &  GW191204\_171526 &  $9.66 ^{+0.38}_{-0.27 }$ &   $0.65^{+0.19}_{-0.25}   $ & \cite{abbott21} & $9.37  \pm{0.44 }$ &  $8.18 \pm{0.31}$ & 0.49$^{+0.14}_{-0.19}$ \\
71  &  GW191215\_223052 &  $24.84^{+2.2 }_{-1.7  }$ &   $1.93^{+0.89}_{-0.86}   $ & \cite{abbott21} & $23.02 \pm{2.71 }$ &  $16.32\pm{3.20}$ & 0.96$^{+0.44}_{-0.43}$ \\
72  &  GW191216\_213338 &  $8.91 ^{+0.22}_{-0.19 }$ &   $0.34^{+0.12}_{-0.13}   $ & \cite{abbott21} & $8.06  \pm{0.29 }$ &  $7.18 \pm{0.28}$ & 0.24$^{+0.08}_{-0.09}$ \\
73  &  GW191219\_163120 &  $4.80 ^{+0.22}_{-0.17 }$ &   $0.55^{+0.25}_{-0.16}   $ & \cite{abbott21} & $4.52  \pm{0.10 }$ &  $3.85 \pm{0.08}$ & 0.38$^{+0.17}_{-0.11}$ \\
74  &  GW191222\_033537 &  $51.04^{+7.1 }_{-5.0  }$ &   $3.0 ^{+1.7 }_{-1.7 }   $ & \cite{abbott21} & $50.98 \pm{0.75 }$ &  $35.78\pm{2.84}$ & 1.66$^{+0.94}_{-0.94}$ \\
75  &  GW191230\_180458 &  $61.68^{+8.2 }_{-5.6  }$ &   $4.3 ^{+2.1 }_{-1.9 }   $ & \cite{abbott21} & $61.26 \pm{0.45 }$ &  $41.72\pm{2.50}$ & 2.24$^{+1.09}_{-0.99}$ \\
76  &  GW200105\_162426 &  $3.62 ^{+0.08}_{-0.08 }$ &   $0.27^{+0.12}_{-0.11}   $ & \cite{abbott21} & $3.65  \pm{0.39 }$ &  $3.35 \pm{0.42}$ & 0.24$^{+0.10}_{-0.10}$ \\
77  &  GW200112\_155838 &  $33.98^{+2.6 }_{-2.6  }$ &   $1.25^{+0.43}_{-0.46}   $ & \cite{abbott21} & $34.73 \pm{1.35 }$ &  $27.05\pm{2.10}$ & 0.85$^{+0.29}_{-0.31}$ \\
78  &  GW200115\_042309 &  $2.58 ^{+0.05}_{-0.07 }$ &   $0.27^{+0.12}_{-0.11}   $ & \cite{abbott21} & $2.58  \pm{0.02 }$ &  $2.53 \pm{0.02}$ & 0.28$^{+0.14}_{-0.10}$ \\
79  &  GW200128\_022011 &  $49.92^{+7.5 }_{-5.5  }$ &   $3.4 ^{+2.1 }_{-1.9 }   $ & \cite{abbott21} & $39.40 \pm{3.80 }$ &  $28.59\pm{1.30}$ & 1.34$^{+0.83}_{-0.71}$ \\
80  &  GW200129\_065458 &  $32.10^{+2/1 }_{-2.3  }$ &   $0.90^{+0.29}_{-0.38}   $ & \cite{abbott21} & $31.55 \pm{2.55 }$ &  $23.28\pm{3.07}$ & 0.53$^{+0.17}_{-0.22}$ \\
81  &  GW200202\_154313 &  $8.16 ^{+0.24}_{-0.20 }$ &   $0.41^{+0.15}_{-0.16}   $ & \cite{abbott21} & $7.42  \pm{0.40 }$ &  $6.46 \pm{0.42}$ & 0.28$^{+0.10}_{-0.11}$ \\
82  &  GW200208\_130117 &  $38.78^{+3.6 }_{-3.1  }$ &   $2.23^{+1.00}_{-0.85}   $ & \cite{abbott21} & $37.18 \pm{3.68 }$ &  $26.65\pm{4.61}$ & 1.14$^{+0.51}_{-0.44}$ \\
83  &  GW200208\_222617 &  $32.54^{+10.7}_{-5.1  }$ &   $4.1 ^{+4.4 }_{-1.9 }   $ & \cite{abbott21} & $33.74 \pm{1.88 }$ &  $25.01\pm{1.30}$ & 2.65$^{+2.83}_{-1.23}$ \\
\hline
\end{tabular}
\end{adjustbox}
\end{table}

\clearpage
\setcounter{table}{0}
\begin{table}
 \caption{~(continuation)}
 \medskip
\begin{adjustbox}{width=\columnwidth,center}
\medskip
\begin{tabular}{ccrrcccc}
  \hline
N   &  ID  &  $\mathcal{M}_\mathrm{GWTC}$ & $D_L^\mathrm{GWTC}$ & {\scriptsize Reference} & $\mathcal{M}_\mathrm{PN}$ & $\mathcal{M}_\mathrm{PN}^\mathrm{c}$ &  $D_L^\mathrm{c}$ \\
\noalign{\smallskip}\hline
\hline
\tiny{1}~ & \tiny{2}~~~~ & \tiny{3}~~~~ & \tiny{4}~~~~~~ & \tiny{5}~~ & \tiny{6}~~~~ & \tiny{7}~~ & \tiny{8}~~ \\
\noalign{\smallskip}\hline
\noalign{\smallskip}
84  &  GW200209\_085452 &  $41.92^{+6.0 }_{-4.2  }$ &   $3.4 ^{+1.9 }_{-1.8 }   $ & \cite{abbott21} & $30.46 \pm{4.54 }$ &  $18.58\pm{3.94}$ & 0.88$^{+0.49}_{-0.46}$ \\
85  &  GW200210\_092254 &  $7.82 ^{+0.38}_{-0.40 }$ &   $0.94^{+0.43}_{-0.34}   $ & \cite{abbott21} & $7.21  \pm{0.34 }$ &  $6.34 \pm{0.42}$ & 0.66$^{+0.30}_{-0.24}$ \\
86  &  GW200216\_220804 &  $53.63^{+9.3 }_{-8.5  }$ &   $3.8 ^{+3.0 }_{-2.0 }   $ & \cite{abbott21} & $48.11 \pm{3.53 }$ &  $26.69\pm{4.71}$ & 1.19$^{+0.94}_{-0.62}$ \\
87  &  GW200219\_094415 &  $43.33^{+5.6 }_{-3.8  }$ &   $3.4 ^{+1.7 }_{-1.5 }   $ & \cite{abbott21} & $43.66 \pm{1.32 }$ &  $27.71\pm{3.21}$ & 1.62$^{+0.81}_{-0.71}$ \\
88  &  GW200220\_061928 &  $117.8^{+23  }_{-15   }$ &   $6.0 ^{+4.8 }_{-3.1 }   $ & \cite{abbott21} & $111.16\pm{5.51 }$ &  $67.82\pm{8.75}$ & 2.39$^{+1.91}_{-1.24}$ \\
89  &  GW200220\_124850 &  $46.81^{+7.3 }_{-5.1  }$ &   $4.0 ^{+2.8 }_{-2.2 }   $ & \cite{abbott21} & $46.78 \pm{1.45 }$ &  $29.67\pm{1.90}$ & 1.87$^{+1.31}_{-1.03}$ \\
90  &  GW200224\_222234 &  $41.05^{+3.2 }_{-2.6  }$ &   $1.71^{+0.49}_{-0.64}   $ & \cite{abbott21} & $39.33 \pm{1.17 }$ &  $27.40\pm{1.60}$ & 0.87$^{+0.25}_{-0.33}$ \\
91  &  GW200225\_060421 &  $17.32^{+1.5 }_{-1.4 }$ &   $1.15^{+0.51}_{-0.53}   $ & \cite{abbott21} & $17.62 \pm{0.87 }$ &  $14.57\pm{0.90}$ & 0.86$^{+0.40}_{-0.40}$ \\
92  &  GW200302\_015811 &  $29.95^{+4.7 }_{-3.0 }$ &   $1.48^{+1.02}_{-0.70}   $ & \cite{abbott21} & $30.78 \pm{2.78 }$ &  $23.11\pm{2.27}$ & 0.96$^{+0.66}_{-0.45}$ \\
93  &  GW200306\_093714 &  $24.15^{+3.5 }_{-3.0 }$ &   $2.1 ^{+1.7 }_{-1.1 }   $ & \cite{abbott21} & $23.46 \pm{1.46 }$ &  $18.42\pm{2.03}$ & 1.33$^{+1.08}_{-0.70}$ \\
94  &  GW200308\_173609 &  $34.77^{+4.8 }_{-2.8 }$ &   $5.4 ^{+2.7 }_{-2.6 }   $ & \cite{abbott21} & $34.16 \pm{1.19 }$ &  $27.21\pm{1.31}$ & 3.58$^{+1.79}_{-1.72}$ \\
95  &  GW200311\_115853 &  $32.72^{+2.4 }_{-2.0 }$ &   $1.17^{+0.28}_{-0.40}   $ & \cite{abbott21} & $33.02 \pm{1.96 }$ &  $25.57\pm{2.69}$ & 0.78$^{+0.18}_{-0.26}$ \\
96  &  GW200316\_215756 &  $10.68^{+0.62}_{-0.65}$ &   $1.12^{+0.47}_{-0.44}   $ & \cite{abbott21} & $9.20  \pm{0.48 }$ &  $8.16 \pm{0.65}$ & 0.71$^{+0.30}_{-0.28}$ \\
97  &  GW200322\_091133 &  $24.80^{+15.7}_{-3.7 }$ &   $3.6 ^{+7.0 }_{-2.0 }   $ & \cite{abbott21} & $23.03 \pm{1.17 }$ &  $16.94\pm{1.63}$ & 1,91$^{+3.71}_{-1.06}$ \\
\hline
\end{tabular}
\end{adjustbox}
\end{table}

\section*{Declarations}

\subsection*{Funding}

The work was partially carried out within the framework of the
state assignment of the Special Astrophysical Observatory,  Russian Academy of Science, 
approved by the Ministry of Science and Higher Education of the Russian Federation.

\subsection*{Competing interests}

All authors certify that they have no affiliations with or involvement in any organization 
or entity with any financial interest or non-financial interest in the subject matter 
or materials discussed in this manuscript.

\subsection*{Ethics approval}

This research does not involve human participants and/or animals.

\subsection*{Data availability}

The data underlying this article are available in
the Gravitational Wave Open Science Center repository, at 
\url{https://doi.org/10.7935/82H3-HH23},~in~the~GitHub 
repository at \url{https://github.com/gwastro/2-ogc},
\url{https://doi.org/10.7935/qf3a-3z67},
~and 
in the public domain resource {\it Stellarcollapse} at \url{https://stellarcollapse.org/bhmasses}.

\end{document}